\documentclass[sigconf]{acmart}

\usepackage{url}
\usepackage{graphicx}
\usepackage{balance}  

\usepackage{xcolor}

\usepackage{algorithm}
\usepackage[noend]{algpseudocode}
\usepackage{array}
\usepackage{subcaption}

\usepackage{mathtools}

\newcommand{\appropto}{\mathrel{\vcenter{
  \offinterlineskip\halign{\hfil$##$\cr
    \propto\cr\noalign{\kern2pt}\sim\cr\noalign{\kern-2pt}}}}}
    
\algnewcommand{\Inputs}[1]{%
  \State \textbf{Inputs:}
  \Statex \hspace*{\algorithmicindent}\parbox[t]{.8\linewidth}{\raggedright #1}
}
\algnewcommand{\Initialize}[1]{%
  \State \textbf{Initialize:}
  \Statex \hspace*{\algorithmicindent}\parbox[t]{.8\linewidth}{\raggedright #1}
}

\AtBeginDocument{%
  \providecommand\BibTeX{{%
    \normalfont B\kern-0.5em{\scshape i\kern-0.25em b}\kern-0.8em\TeX}}}

\begin{document}

\copyrightyear{2021}
\acmYear{2021}
\setcopyright{acmcopyright}
\acmConference[SIGMOD '21]{2021 International Conference on Management of Data}{June 20-June 25, 2021}{Xi'an, Shaanxi, China}
\acmBooktitle{2021 International Conference on Management of Data (SIGMOD '21), June 20-June 25, 2021, Xi'an, Shaanxi, China}
\acmPrice{15.00}
\acmDOI{10.1145/XXXXXXX.XXXXXXX}
\acmISBN{978-1-4503-XXXX-X/21/06}

\title{Scaling Strongly Consistent Replication}


\author{Aleksey Charapko}
\authornote{Work partially completed at the University at Buffalo, SUNY.}
\affiliation{University of New Hampshire}
\email{Aleksey.Charapko@unh.edu}

\author{Ailidani Ailijiang}
\affiliation{Microsoft}
\email{Ailidani.Ailijiang@microsoft.com}

\author{Murat Demirbas}
\affiliation{University at Buffalo, SUNY}
\email{demirbas@buffalo.edu}

\begin{abstract}

Strong consistency replication helps keep application logic simple and provides significant benefits for correctness and manageability. Unfortunately, the adoption of strongly-consistent replication protocols has been curbed due to their limited scalability and performance. To alleviate the leader bottleneck in strongly-consistent replication protocols, we introduce Pig, an in-protocol communication aggregation and piggybacking technique. Pig employs randomly selected nodes from follower subgroups to relay the leader's message to the rest of the followers in the subgroup, and to perform in-network aggregation of acknowledgments back from these followers. By randomly alternating the relay nodes across replication operations, Pig shields the relay nodes as well as the leader from becoming hotspots and improves throughput scalability.

We showcase Pig in the context of classical Paxos protocols employed for strongly consistent replication by many cloud computing services and databases. We implement and evaluate PigPaxos, in comparison to Paxos and EPaxos protocols under various workloads over clusters of size 5 to 25 nodes. We show that the aggregation at the relay has little latency overhead, and PigPaxos can provide more than 3 folds improved throughput over Paxos and EPaxos with little latency deterioration. We support our experimental observations with the analytical modeling of the bottlenecks and show that the rotating of the relay nodes provides the most benefit for reducing the bottlenecks and that the throughput is maximized when employing only 1 randomly rotating relay node.


\end{abstract}

\maketitle

\section{Introduction}

Strong consistency replication ensures that data viewed immediately after an update will be consistent for all observers of the entity~\cite{linearizability}. This helps to keep application logic simple and provides significant benefits for correctness and manageability. Without strong consistency, the developers are forced to write a lot of peripheral code to handle side-effects and corner-cases arising from concurrent eventually-consistent operations. Moreover, an unanticipated combination of faults conspiring with the concurrency of operations is likely to leave those systems in states that are hard to debug and repair~\cite{taxdc,cloudbugs,jepsen}.

Unfortunately, the adoption of strongly-consistent replication protocols has been curbed due to their limited scalability and performance. In strongly consistent replication, all the communication drains through a single node -- the primary (a.k.a. the leader), which constitutes a throughput bottleneck~\cite{dissecting_perf}. The leader performs a disproportionately large amount of work compared to its $N\!-\!1$ followers. For each consensus instance, the followers receive one message from the leader and send one message back. In contrast, the leader needs to send $N\!-\!1$ messages to the followers, and receive at least a quorum of messages back from these followers, not counting the interaction with clients. For instance, in a cluster of $N=5$ nodes, the leader handles up to 4 times more messages than a follower. This hinders scalability in terms of both adding more nodes and serving more operations -- adding more followers for increased fault tolerance or weaker-consistency scalable reads only increases the load on the leader linearly and reduces its performance~\cite{dissecting_perf}. At the same time, to scale the system for more strongly consistent operations, we need to reduce the disproportionate workload on the leader.

To address the single leader bottleneck and improve the throughput of strongly-consistent replication protocols, we introduce Pig, an in-protocol communication aggregation and piggybacking technique which aims to decouple the decision making from the communication at the leader. 

Pig replaces the direct communication between the leader and followers with a relay-/aggregate-based message flow.
%
In particular, Pig employs randomly selected nodes from one or more follower subgroups to relay the leader's message to the rest of the followers in each subgroup, and to perform an in-network aggregation of responses back from these followers. This overlay communication tree allows the leader to only exchange messages with a small number of relay nodes.
Moreover, the random alternation of the relay nodes across replication operations shields these relay nodes from becoming hotspots themselves: the extra traffic load a relay node incurs in one round is offset in consecutive rounds when the node no longer serves as a relay. These two properties combined improve the throughput scalability of the system.

Although aggregation-based approaches have been known and employed in the context of weak-consistency replication protocols~\cite{optimisticreplication}, Pig demonstrates how they can be effectively applied to and integrated with strong consistency replication protocols.
In particular, our experiments and analysis show that the rotation of the relay nodes in the Pig provides the most benefit for reducing the bottlenecks in strongly-consistent replication and improving throughput.
Another novel and somewhat counterintuitive finding from our experiments and analysis is that the communication bottlenecks are minimized when only one relay node is employed in contrast to what seems to be a more balanced configuration of using $\sqrt{N}$ relay nodes.


In this paper, we showcase Pig in the context of classical Multi-Paxos~\cite{paxos} protocols, which, owing to their excellent fault-tolerance properties, are employed for strongly consistent replication by many cloud computing services and databases~\cite{chubby,zookeeper,paxosLive,configerator,megastore,zab,delos,logdevice,kafka,slog,borg,spanner,bizur,cockroachdb,yugabyte,paxosstore}.
Despite significantly improving scalability and performance, PigPaxos reuses the same correctness proof as Paxos, because it preserves the core protocol, and modifies only the communication implementation to improve scalability and performance of the protocol.
As such, the communication piggybacking and aggregation technique used in Pig is applicable to many Paxos implementations~\cite{raft,etcd} and variants~\cite{fpaxos,wpaxos}.

Moreover, PigPaxos tolerates up to $f$ node failures in a cluster of size $2f+1$, just like the classical Paxos protocol. However, it can scale to large clusters with up to 25 nodes. Such large clusters may be useful in environments with more frequent failure rates, as to tolerate a greater absolute number of failures. Additionally, many data-management and database systems allow relaxed consistency guarantees. Relaxed consistency models, such as consistent prefix~\cite{terry2013replicated} or session reads like in ZooKeeper~\cite{zookeeper} can benefit from having more replicas in the cluster to scale read throughput.
%



%
We implement PigPaxos in the Paxi framework~\cite{dissecting_perf} and evaluate many optimizations possible over Pig's basic dynamic communication overlay tree scheme. We also employ Paxi as a level playground to compare and benchmark Paxos, EPaxos, and PigPaxos in the context of the same in-memory key-value store implementation.
Our experiments, conducted on AWS EC2 nodes with 5-25 nodes in various LAN and WAN topologies, show that PigPaxos is surprisingly effective in scaling consensus to large clusters. For a 25 node deployment using a uniformly random workload in a 1000 item datastore, EPaxos throughput gets saturated at 3000 requests per second, Paxos throughput reaches its limit of around 2000 req/sec, whereas PigPaxos scales beyond 7000 req/sec with little latency deterioration.

Our experiments also show that the aggregation has little overhead for the latency of PigPaxos as compared to the latency of Paxos. Another interesting result from our evaluation is that PigPaxos provides benefits over Paxos for consensus clusters as small as 5 nodes. In particular, due to aggregation at the relay nodes, PigPaxos throughput scales with respect to the size of the messages. This makes PigPaxos applicable for distributed database systems that use Paxos for replication such as CockroachDB~\cite{cockroachdb} and Spanner~\cite{spanner}. PigPaxos also provides benefits for implementing geo-replicated distributed databases deployed over many regions around the globe, as such planetary-scale systems often have up to a few dozen nodes supporting a single partition~\cite{cosmosdb_global_distr}.

We corroborate our experimental observations with the analytical modeling of the bottlenecks and show the extent to which PigPaxos helps shift these bottlenecks from the leader to the followers. We also provide formulas to estimate the relative load of different PigPaxos components in various configurations. For instance, we demonstrate that the cluster size has a limited and capped impact on follower load, while the number of relay groups significantly affects the leader load.

{\bf Paper outline.} \ Section~\ref{sec:background} provides a summary of Paxos. Section~\ref{sec:PigPaxos} presents the PigPaxos protocol, and Section~\ref{sec:optimizations} presents optimizations. Evaluation results are presented in Section~\ref{sec:eval}, and discussion and detailed analysis of the findings are provided in Section~\ref{sec:disc}. We discuss related work in Section~\ref{sec:related}, followed by the future work and extension in Section~\ref{sec:future}, and conclude in Section~\ref{sec:conclusion}. 

\label{sec:intro}

\section{Background}
\label{sec:background}
\subsection{Paxos Overview}
\begin{figure}[t]
	\centering
	\includegraphics[width=\columnwidth]{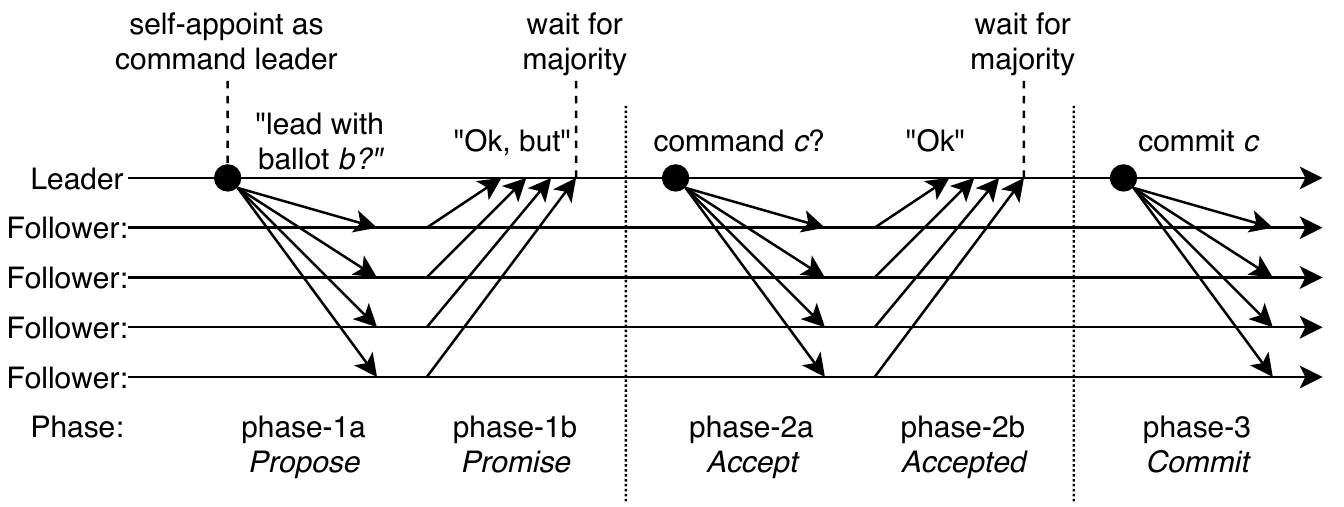}
	\caption{The three phases of Paxos protocol}
	\label{fig:paxos}
\end{figure}

The classical Paxos protocol~\cite{paxos} operates in three phases as illustrated in Figure~\ref{fig:paxos}. The first phase often called the propose phase, establishes a leader node. In this phase, a node tries to acquire leadership by reaching out to other nodes with some ballot number. The replicas ack the leadership proposal only if they have not seen a higher ballot. When a node collects a majority of acks, it can proceed to the second phase. In phase-2, the accept phase, the leader tells all the followers to accept a command. The command is either a new command of the leader's choosing or an old command if some nodes replied during phase-1 with an earlier uncommitted command. Once the leader receives a majority of acks from nodes accepting the command, the command is anchored, and the leader proceeds to the commit phase (phase-3). The leader then instructs the followers to finalize the command in the log and execute it against their state machines provided that there is no gap in their logs up to this command's slot.

\begin{figure}[t]
	\centering
	\includegraphics[width=\columnwidth]{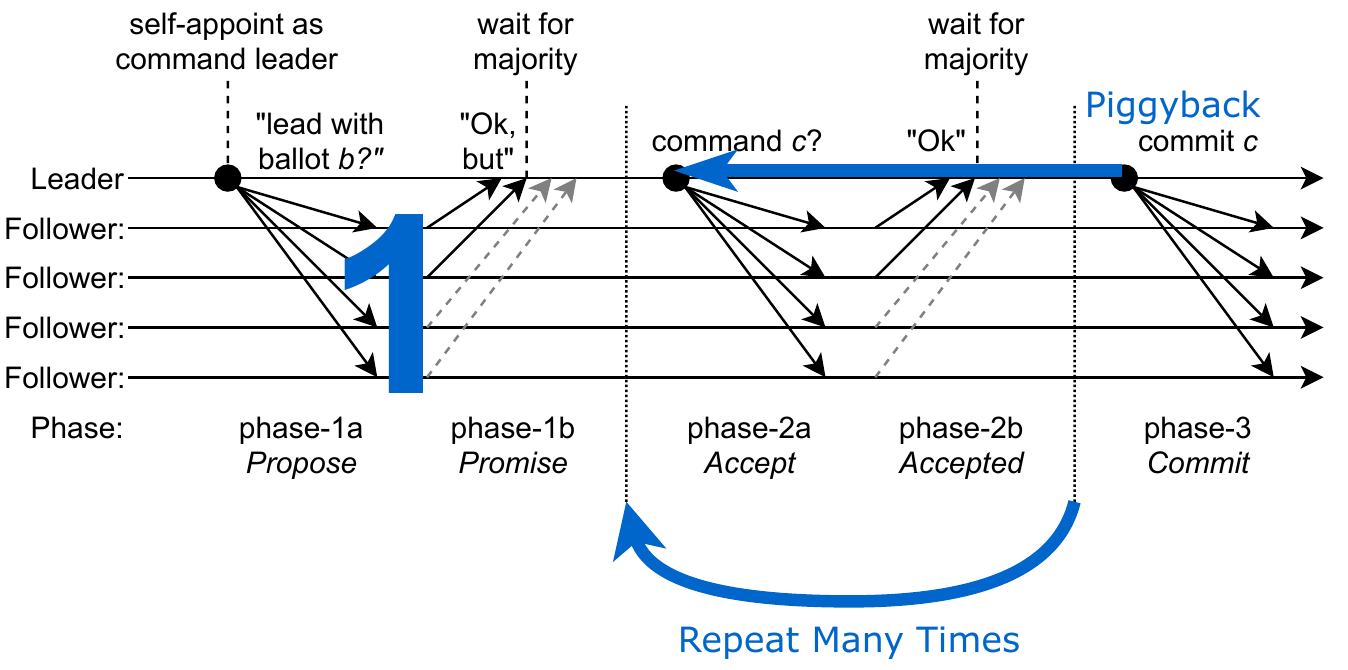}
	\caption{Multi Paxos optimization}
	\label{fig:multipaxos}
\end{figure}

This basic Paxos protocol is commonly extended to the Multi-Paxos~\cite{paxosModerate} protocol, which allows the same leader to propose commands in subsequent slots (i.e., consensus instances) as illustrated in Figure~\ref{fig:multipaxos}. The leader election in phase-1 is performed only once, and this cost is avoided in subsequent consensus instances as long as the incumbent leader's ballot number remains the highest the followers have seen. Additionally, phase-3 is piggybacked to the next phase-2, further reducing the communication overheads. In the rest of the paper, when we mention the Paxos protocol, we refer to this Multi-Paxos optimization.

\subsection{Leader Bottleneck}
Paxos, like many other strongly consistent replication protocols, relies on a strong single leader to coordinate and order the operations in the system. This strong leader, however, is often a bottleneck, especially when every read and write operation has to go through it. The bottleneck has been intuitively known for quite some time, and many databases and storage systems~\cite{cassandra,dynamo} purposefully picked a weaker consistency model to avoid a strong single leader and load balance themselves by shedding the workload equally among all the nodes.
Some strongly consistent replication solutions also take a similar approach. For instance, EPaxos~\cite{epaxos} avoids a single leader through the means of opportunistic, operation-level leadership. In Sections~\ref{sec:eval} \& \ref{sec:related} we provide more insights into EPaxos's performance and operation.
More recently, the intuition for the leader bottleneck was tested experimentally and analytically~\cite{dissecting_perf}, and it was shown that the bottleneck occurs largely because of the extensive communication with the leader. A strong leader needs to send messages to all the replicas, and receive responses to know when the operation has been successfully replicated in the state machine or log.

This fan-in/fan-out communication pattern is demanding for the leader. In the more common case, the leader will be bottlenecked at the CPU serializing, deserializing, and processing these messages. For example, in our testing, we have experimentally verified that serializing and deserializing just around 100,000 phase-2a and phase-2b messages are enough to saturate one CPU core of our test machine. However, the node also needs to handle processing these messages, further consuming limited CPU cycles. It is also worth mentioning that in some rare cases when the messages are very large, the bottleneck may shift towards the available network bandwidth. In both cases, the root cause of the problem is the same -- too many messages are sent, received, and processed by one node.

In Section~\ref{sec:PigPaxos} we introduce the Pig primitive to alleviate the leader communication problem in strongly consistent replication protocols. We then apply Pig to Paxos and overhaul its communication implementation, resulting in a more scalable protocol, PigPaxos. Since Paxos safety and liveness proofs are oblivious to how the communication between the leader and followers is implemented, Pig comes as an orthogonal technique that can be applied in the context of many Paxos variants for improving throughput.

\section{PigPaxos}
\label{sec:PigPaxos} 
\subsection{Pig Primitive}
\label{sec:Pig} 



Strongly consistent replication protocols, such as Paxos and primary-backup protocols, exhibit similar communication patterns. They rely on a single stable leader used for ingesting and ordering the operation. The leader replicates the operations along with their order positions to the follower nodes, and the followers reply to the leader to acknowledge the operations. We can break down this communication into two steps: fan-out message flow from the leader, and fan-in flow to the leader. 

The major problem with both fan-out and fan-in communication flows is that they disproportionally bog down the node initiating the communication -- the node needs to send messages to $N-1$ other nodes, where $N$ is a total number of machines/processes, and then receive $N-1$ distinct replies. This communication pattern is the source of the leader-bottleneck in replicated state machines~\cite{dissecting_perf}. Pig is a simple primitive designed to solve this problem and improve the efficiency of fan-out and fan-in message flows. Pig builds on top of a multicast idea to disseminate the messages from one source to multiple recipients efficiently. Unlike multicast, Pig assumes two-way communication and aggregates the responses coming back from the recipients.

Pig eschews direct communication between the initiating and replying nodes. Instead, the communication initiator picks another random node as the relay and sends the message to it. The job of a relay is threefold. Firstly, it acts as a regular node in the system and processes the incoming message. Secondly, it re-transmits the message over to the remaining nodes. Thirdly, it acts as an aggregator for the replies on the way back to the initiator.

\begin{figure}[t]
	\centering
	\includegraphics[width=\columnwidth]{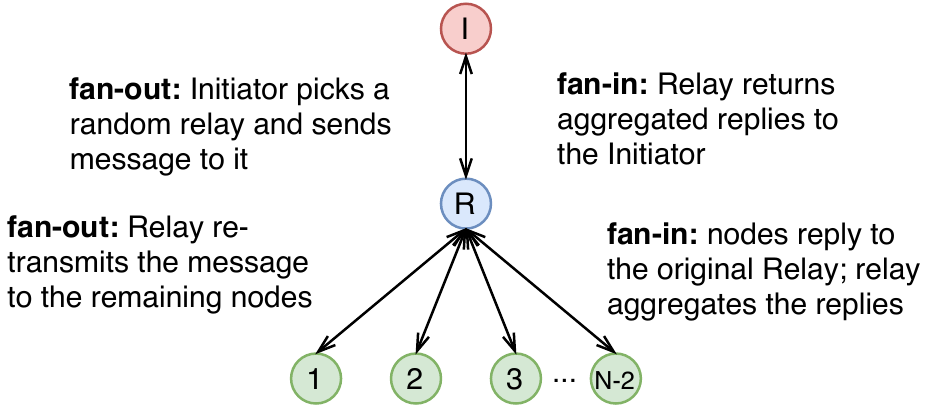}
	\caption{Pig communication pattern for a single fan-out \& fan-in pair.}
	\label{fig:pig_communication}
\end{figure}

The later aspect of Pig is the major difference from a peer-to-peer multicast. When nodes reply to the message, they reply to the relay instead of the communication initiator. The relay is then aggregating the replies together and piggybacks them into one concise message to the initiator. We illustrate the Pig communication pattern in Figure~\ref{fig:pig_communication}


At first glance, Pig may appear as inefficient as a regular fan-out/fan-in round trip, since now a single relay node is doing all the heavy work instead of the initiator. However, as we repeat multiple Pig rounds, we pick the relay nodes uniformly randomly, amortizing the extra work done by the relay over multiple rounds in which a node does not serve as a relay.  This approach works well when consecutive the fan-out/fan-in communication rounds are not purely sequential and multiple concurrent communication rounds can take place. This works similar to pipelining -- a node may serve as a relay for the communication round $i$, and as a follower for other largely concurrent rounds $i+1, ..., i+n$.

Our implementation of Pig hides behind two general APIs: \\ {\tt broadcast(broadcastMsg)} and {\tt send(sendMsg, toNode)}. The
{\tt broadcast(broadcastMsg)} API sends the message {\tt broadcastMsg} from the initiator to all the nodes participating in the current message exchange round. The {\tt broadcastMsg} must contain a {\tt PigMsgID} to uniquely identify the Pig round. Every node, including the relay, replies to the relay node using the {\tt send(sendMsg, toNode)} API. When replying, {\tt sendMsg} must contain the same {\tt PigMsgID} as the original fan-out message. The relay uses this {\tt PigMsgID} to match the reply with the original fan-out message and aggregate all relevant replies together. Once the relay receives the required threshold of replies (or it times out), it forwards the aggregated reply to the initiator, completing the Pig communication pattern. Our {\tt send(sendMsg, toNode)} API is general enough such that if the {\tt sendMsg} is not part of any Pig instance, then the  {\tt sendMsg} gets delivered as a regular point-to-point message without any Pig relaying or aggregation. 


\label{sec:pig}

\subsection{Putting Pig into Paxos}

\begin{figure}[t]
	\centering
	\includegraphics[width=\columnwidth]{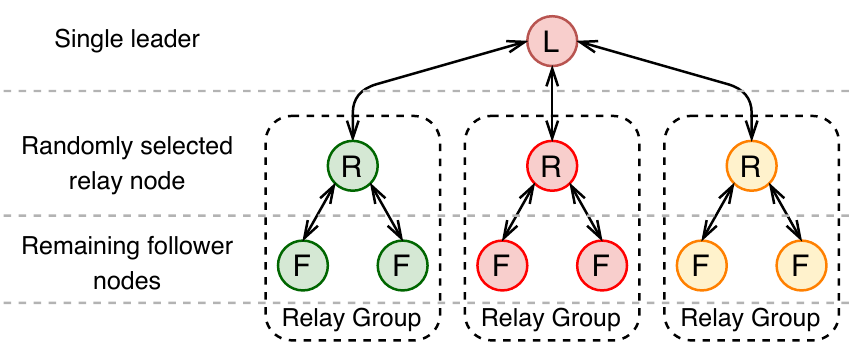}
	\caption{PigPaxos communication flow. All nodes are divided into groups, and the leader communicates only with a random node from each group. The nodes contacted by the leader relay the messages to their peers within the group and aggregate their responses to reply back to the leader.}
	\label{fig:pigpaxos_communication_relay}
\end{figure}


Leveraging our observation on common message exchange patterns in Paxos, we can easily apply our Pig communication flow to Paxos-family of replicated state machines. We can use one or more instances of Pig in each Paxos deployment to shard both the fan-in and fan-out communication across the nodes. Figure~\ref{fig:pigpaxos_communication_relay} illustrates how multiple Pig instances can be overlaid in the Paxos cluster. Initially, all nodes are divided into several distinct non-overlapping relay groups. This grouping may happen with the help of a hash function or maybe pre-defined in terms of the node topology. For example, in a geo-distributed setup, we may statically pre-define the groups based on the regions or datacenters in which nodes are located. Our reference implementation uses such static assignment of nodes to groups. Within each static group, we use Pig for communication from the leader to the followers in the group.
%

\begin{figure}[t]
	\centering
	\includegraphics[width=\columnwidth]{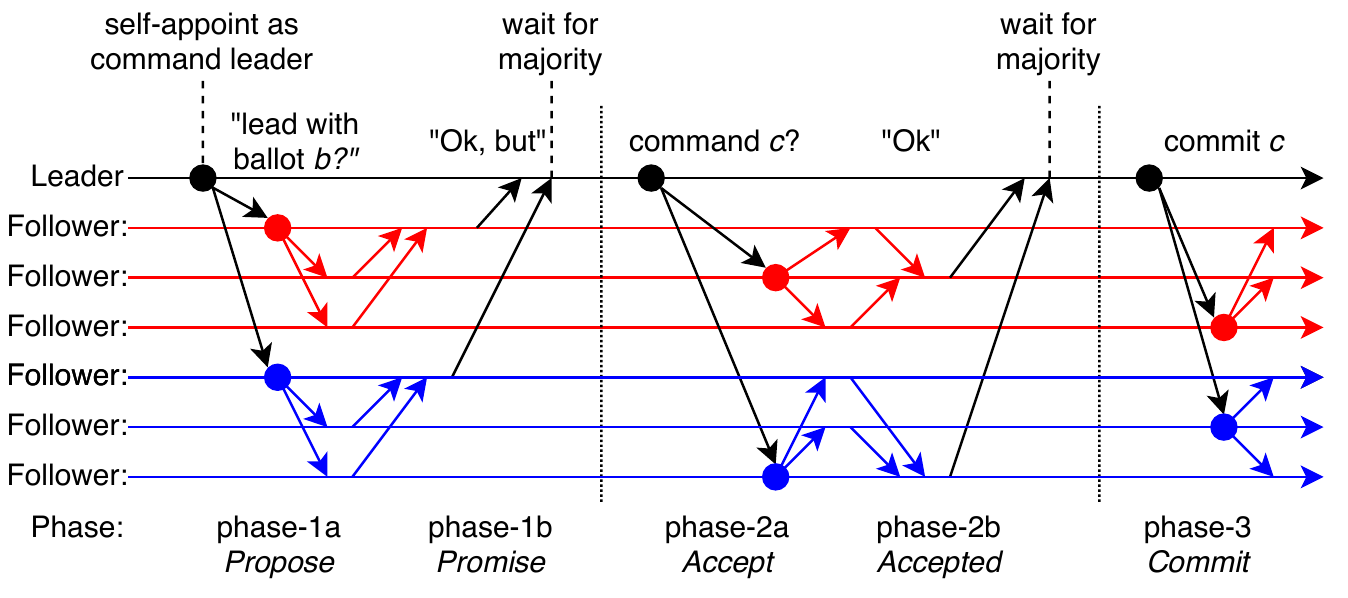}
	\caption{The three phases of PigPaxos with a single level communication relay.}
	\label{fig:bigpaxos}
\end{figure}

In Figure~\ref{fig:bigpaxos} we demonstrate the Pig communication flow of PigPaxos for Phase-1, Phase-2, and Phase-3. Note that, using Multi-Paxos optimization with a stable leader, PigPaxos would only perform Phase-2 for each consensus instance, with Phase-3 messages piggybacked to Phase-2. We break-down and describe the communication flow consisting of a fan-out and fan-in as follows.
\begin{enumerate}
\item When the leader starts a fan-out communication, it starts the Pig pattern for each relay group, by picking a random relay from every group. The random rotation of relay nodes is important for load-balancing the communication burden to all followers and avoiding hotspots. As relay nodes alternate, the extra load a relay node suffers in one round is offset in the other concurrent, but consecutively numbered,  rounds in which that node ceases to be a relay.

\item Upon receiving a message from the leader, a relay node processes this message as a regular follower and resends the message to the remaining nodes in its relay group.

\item Upon receiving the messages, the followers start the fan-in pattern by responding back to the relay node as if they were responding directly to the leader.

\item Each relay node waits for the followers' responses and piggybacks them together into a single message. By default, the relays wait for all followers in the group to respond. Since such unbounded wait may create performance problems when some follower becomes sluggish or fail, PigPaxos employs a tight timeout at the relay nodes. If some followers do not reply within the timeout, a relay stops waiting and replies to the leader with all responses collected so far.

\item The leader performs final aggregation of replies from each relay group and decide whether the quorum of votes have been received.
\end{enumerate}

In some cases, we can break the communication pattern short, when it is obvious that no further progress can be made. Usually, these short-cuts arise during a leader change. For example, when a relay node learns of a new leader with a higher ballot either through a direct message or a reply from a follower, it no longer tries to finish the aggregation, and replies to the old leader with a reject message carrying the new ballot.

Since PigPaxos uses several instances of Pig for each Paxos round, it can adjust to a variety of different situations. In geo-distributed setup, it can be beneficial to locate groups in regions to minimize the WAN traffic. In local area networks, the same trick can apply within the availability zones.  Additionally, a different number of relay groups may have an impact on performance under failures -- one relay failure causes the corresponding relay group to not return the data on time. This may potentially slow down the protocol if the remaining health relay groups are not enough to form the majority. We discuss the fault tolerance and liveness further in Section~\ref{sec:ft}.

\subsection{Paxos to PigPaxos mapping}
PigPaxos generalizes the communication implementation of the Paxos protocol. Paxos has $N-1$ groups, where each group has one element and groups do not intersect with each other. In contrast in PigPaxos there are $p$ groups where $p \in \{ 1..N-1\}$, and the cardinality of the union of $p$ groups is $N-1$. This formulation allows the subgroups to intersect with each other, which may have advantages in terms of adding redundant channels to reach some nodes. However, for simplicity's sake, we preclude intersecting groups in the rest of the paper.

We note that the safety proofs of Paxos~\cite{paxos} do not depend on the communication implementation between the leader and follower nodes.
In Paxos, maintaining correctness in spite of failures is guaranteed by quorum size and the information exchanged within the quorums, and the proofs are oblivious to how the communication of this information is achieved.
Therefore, PigPaxos preserves the safety and liveness properties of Paxos, as it only modifies the communication implementation.
For reasoning about liveness, the message flow pattern and the use of relay nodes requires special attention, as a failure of a relay node has a disproportional impact compared to the failure of a regular follower. Similar to Paxos, PigPaxos requires some level of synchrony to make progress in a fault-tolerant manner, as FLP impossibility prevents consensus in fully asynchronous networks with machine failures~\cite{flp}. PigPaxos assumes some upper bound delay on message propagation and retries if communication takes longer. This ensures that when the synchrony is restored to the adequate levels as prescribed by the upper bound delay, the protocol will make progress. 
Since PigPaxos uses a uniformly random selection of relay/aggregator nodes at each round, it circumvents the problem of having some minority failures denying all progress. 

\subsection{Fault Tolerance}
\label{sec:ft}
Similar to Paxos, PigPaxos considers a crash failure model in which nodes may silently crash.
PigPaxos tolerates up to $f$ node failures in a cluster of size $2f+1$, just like the classical Paxos protocol, since it implements an unchanged Paxos algorithm.
However, Pig communication flow introduces its failures into the PigPaxos, potentially causing slightly different performance behavior than classical Paxos implementation, as failures can cause the system to retry Pig instances, delaying the message propagation. We distinguish two common communication failure cases: a follower failure and a relay node failure. 

\begin{figure}[!t]
\centering
\begin{subfigure}{0.35\textwidth}
\includegraphics[width=\linewidth]{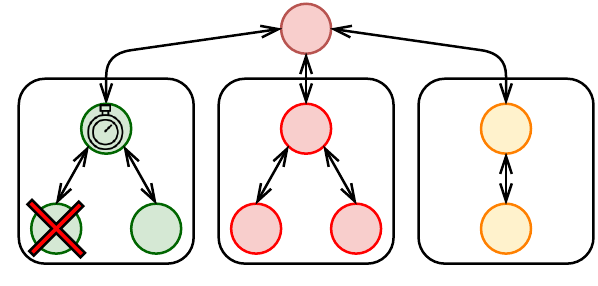}
\caption{Follower failure causes the relay node to timeout before responding to the leader.}
\label{fig:pigpaxos_leaf_failure}
\end{subfigure}

\begin{subfigure}{0.35\textwidth}
\includegraphics[width=\linewidth]{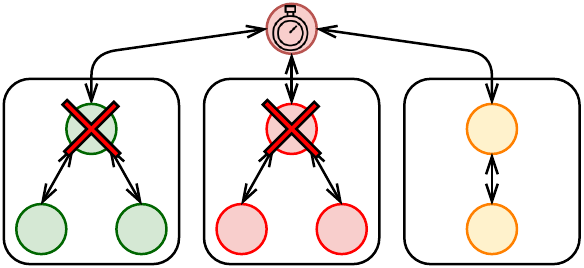}
\caption{Relay node failures may break message delivery, causing leader to timeout and retrying with different relays. }
\label{fig:pigpaxos_relay_failure}
\end{subfigure}
\caption{Failure scenarios at follower and relay nodes.}
\label{fig:failures}
\end{figure}

PigPaxos deals with a follower failure by enacting a tight timeout on the relay node -- the relay node does not wait for responses indefinitely and will reply to the leader either upon collection of all children responses or reaching a timeout $T_r$. The relay timeout procedure allows a partial set of responses to reach the leader in the hopes that a majority is still satisfied. We illustrate this type of failure in Figure~\ref{fig:pigpaxos_leaf_failure}. In many cases, the leader remains unaffected by the timeout of a minority of relay groups, as the majority quorum may be reached by votes from other relay groups.


Relay node failures are more severe. When a relay fails, the leader will receive no feedback/votes from that relay group at all. If a sufficient number of votes are tallied with other relay groups, the protocol will proceed as normal, however, if the leader cannot get the majority quorum, it will timeout. Figure~\ref{fig:pigpaxos_relay_failure} demonstrates this failure scenario. 

Regardless of the failure type, a healthy PigPaxos leader institutes a retry upon reaching the timeout. A retry, just like any Pig round will use a new, randomly selected relay group to make sure that a few specific failures cannot bring the system down. Upon every retry, we draw the new relay nodes from a uniformly random distribution, giving no preference to any node. Under severe failures when close to $f$ nodes fail, the retries may become more frequent, since it may take multiple attempts to randomly find a Pig instance with healthy relays to collect enough responses. However, this does not cause availability loss, as the system remains live and will make progress, albeit at a reduced rate.
%
%

PigPaxos has two distinct timeouts -- one for the relay nodes and one for the leader. These timeouts have a strict relation, as the relay node timeout needs to be much smaller than the leader timeout to allow the relays to deliver the partial response to the leader before the leader reaches its timeout.
The timeout retries do not affect the safety of the protocol, as they only hinder message delivery and do not affect decision making at the leader node. Retrying a phase-2 prepare message cannot cause any problem in Pig, since the leader keeps track of all nodes that have voted on the slot in a given Paxos ballot. If and when a duplicate vote arrives, the leader will recognize it and not count  the vote from an already seen node twice. 
Furthermore, the leadership itself is protected by the ballot number, allowing a different node to use regular Paxos means of overtaking the leadership as needed.
As a result, if some new leader is in a better position to communicate with the cluster, it can get elected with a higher ballot and proceed, while resend attempts by the old leader will fail as having an obsolete ballot.



\section{PigPaxos Optimizations}
\label{sec:optimizations}

\subsection{Partial Response Collection}
\label{sec:partil_response_collection}
In the basic PigPaxos protocol, the relay nodes wait to hear from all nodes in
their relay groups before they aggregate and forward these messages to the leader.\footnotemark~ If some nodes fail to respond, the relay nodes time-out and send whatever messages were collected to the leader.
This may cause slowdowns when some nodes in the groups are sluggish or crashed.
One way to mitigate the problem is to avoid waiting for all followers in the group to reply, and send a partial response to the leader when sone threshold $g_i$ of nodes in a group $i$ have responded. We define $g_i$ in terms of the number of nodes ($PRC$) we are willing to omit in the aggregation: $g_i=n_i-PRC$, where $n_i$ is the size of the group $i$. It is still important that the leader collects at least a majority
votes across all relay groups, so the threshold $g_i$ (or rather the number of slacking nodes $PRC$), must be chosen accordingly:
$\sum_{i=1}^{R}g_i >= \lfloor\frac{N}{2}\rfloor + 1$, where $R$ is number of
relay groups and $N$ is the total number of nodes.
This will improve the per-round latency of PigPaxos as well as reduce momentary slowdowns due to some sluggish or failed nodes. However, this may also cause issues with more severe failures -- a crashed relay group and low $g_i$ (high $PRC$) may cause the leader to not receive enough replies, requiring more communication to learn the missing votes.

\footnotetext{
Of course, if a relay node receives a reject message (either at Phase-1 or Phase-2) it does not wait for aggregation and sends this rejection to the leader immediately.}

\subsection{Gray List of Potential Failures}
\label{sec:gray}
Selecting relay nodes from a uniformly random distribution is an easy strategy that provides liveness even under severe failures with enough retries. However, even a small number of failures may negatively impact the latency of the system in some configurations. For example, consider a two relay groups setup in $2f+1$ cluster. One group will consist of $f+1$ nodes, while the other group is a minority one with just $f$ machines. In such a configuration, even a single failure in the majority group can bring adverse performance effects -- when a failed node is selected as a relay, it will cause the entire relay group to time-out, increasing the operation's latency.

A simple solution to combat the problem is to keep track of such failed, slow, or potentially problematic nodes. Each node maintains a gray list of potentially problematic nodes, such as nodes that have had network failure. The leader, upon selecting a random relay will exclude the gray-listed nodes from the set of potential relays for some time. However, gray-listed nodes may still be picked as relays on occasion as a way to probe their health. 

\subsection{Special Case of Single Relay Group}
PigPaxos does not restrict the number of relay groups. As we showed earlier, when $R=N-1$, PigPaxos becomes a normal Multi-Paxos system. Another special relay group configuration is $R=1$. In this setup, the leader needs to communicate with just a single relay node, potentially giving the best possible throughput by removing as much work from the leader as possible. However, waiting for all nodes to reply to the relay in $R=1$ becomes problematic in case of slow or failed nodes. Configurations with $R>1$ often mask one slow relay group by getting a majority votes from other groups. Fortunately, a single relay group provides a relay node with nearly-global cluster view, allowing it to use {\em partial response collection} more efficiently. When the majority of nodes have replied to the relay, it can safely respond to the leader without waiting for the minority of slow nodes. This is not possible with more than one relay group, as each relay node has only partial information about all replied nodes, and cannot make a decision on whether the global majority has been reached.


\section{Performance Evaluation}
\label{sec:eval}
\subsection{Implementation \& Testbed}
\begin{figure}[t]
\centering
\includegraphics[width=\linewidth]{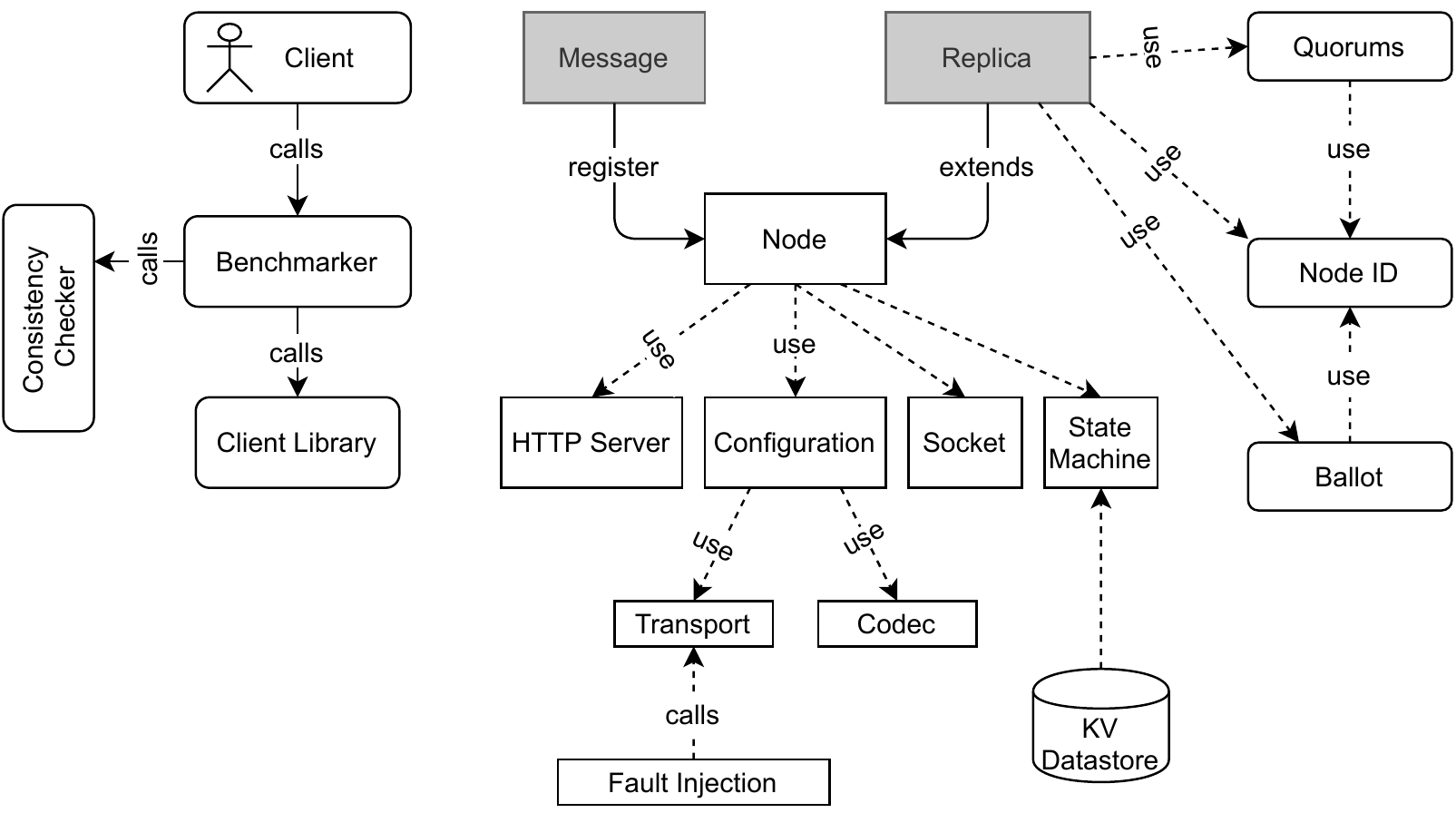}
\caption{\textbf{Paxi architecture redrawn from~\cite{dissecting_perf}. We implemented PigPaxos by overriding the Messages and Replica components (shaded).}}
\label{fig:paxi}
\end{figure}

We implemented PigPaxos in Go using Paxi~\cite{dissecting_perf}.
Paxi is an open-source testbed for prototyping, evaluating, and benchmarking consensus and replication protocols. As shown in Figure~\ref{fig:paxi}, Paxi readily provides most functionality that any coordination protocol needs, including network communication, the state machine of a key-value store, and multiple types of quorum systems. Paxi also includes a built-in YCSB-like~\cite{ycsb} benchmarking utility for testing of all protocols implemented in a framework.

Our implementation of PigPaxos in the Paxi framework required almost no changes to the core Paxos code, and focused only on the message-passing layer and relay group orchestration. The entire protocol was implemented in just 1200 lines of code, and is available as opensource at \url{https://github.com/pigpaxos/pigpaxos}. The performance of Paxi's Paxos implementation is on par with industry implementations~\cite{dissecting_perf}.

We conducted our experiments on AWS EC2 m5a nodes. All Paxos and PigPaxos nodes used m5a.large instances with 2 vCPUs. To ensure that the client performance does not impact the results, we used the larger m5a.xlarge instances with 4 vCPUs for the clients.

We use closed-loop or synchronous clients, meaning that each client can have only one outstanding operation and will not start a new operation before the completion of the previous one. To drive the throughput in the system we used up to 120 clients, better approximating a real environment with many clients interacting with the system. We measure latency at the client, meaning that the operation latency is the time between issuing a request at the client and receiving a successful response.

The replication protocols implemented in Paxi use an internal in-memory key-value store for testing purposes. Paxi's benchmark utility is capable of generating YCSB workloads against this internal datastore and provide a fair comparison as the only components that change under test are replication protocols. For our Multi-Paxos and PigPaxos evaluation, we used a workload with 1000 distinct keys and a uniform probability of selecting each key. It is worth noting that systems driven by a single Multi-Paxos instance do not experience any performance variation due to the workload skew, since all operations are placed in the same single write-ahead log. However, leaderless protocols, such as EPaxos~\cite{epaxos} do experience conflicts due to the workload skewness. Using 1000 keys allowed us to adjust the workload for EPaxos to avoid any conflicts and present the best possible leaderless scenario. 

We used a KV-pair with small keys and values of 8 bytes in most tests, however, in select experiments we increased the size of the value up to 1.25 Kb. We used an even distribution of reads and writes, since the generic implementation of Paxos does not differentiate between different operation types, and both reads and writes must go through the replication (phase-2) of Paxos. Unless otherwise stated, we used a partial response collection strategy and did not wait for the slowest node in each group.

\subsection{Number of Relay Groups}
\label{sec:num_relays}
\begin{figure}[t]
	\centering
	\includegraphics[width=\columnwidth]{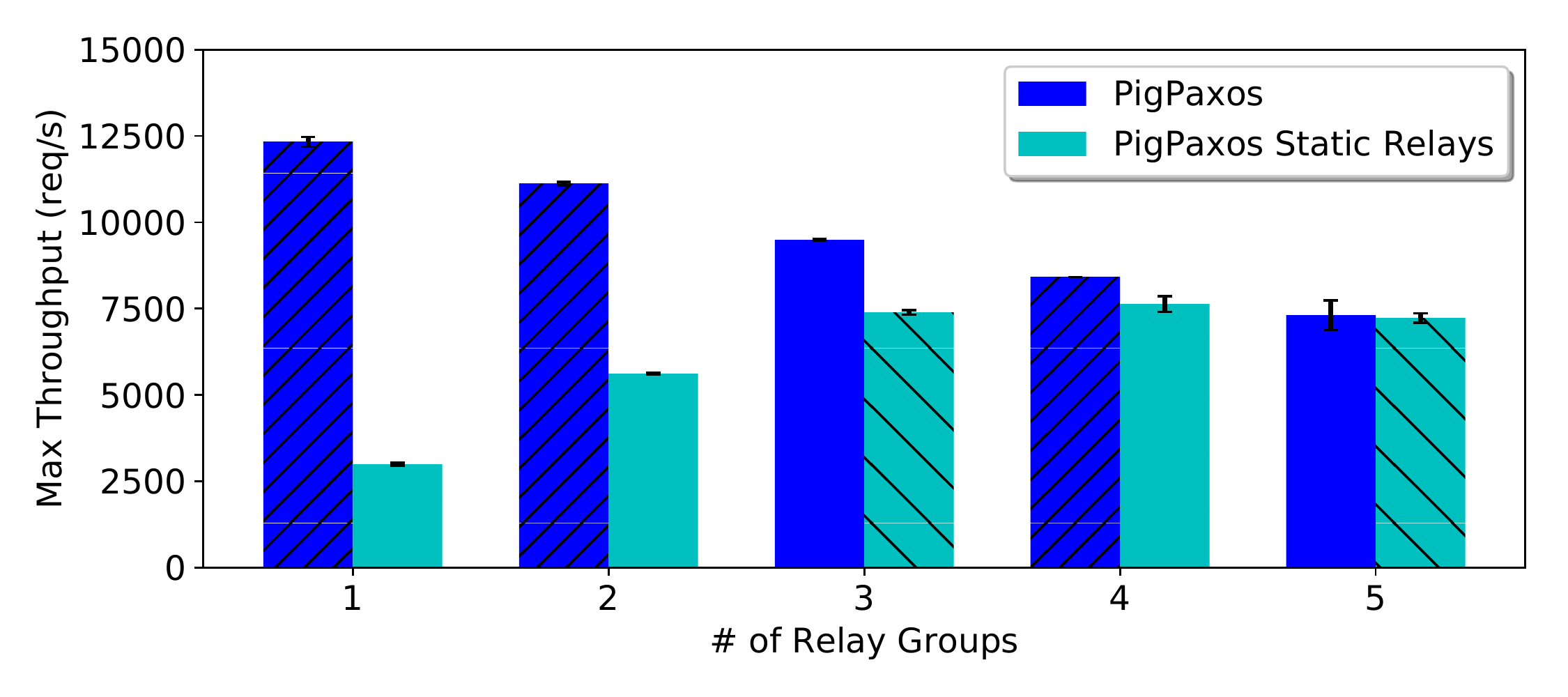}
	\caption{Number of relay groups on 25-node PigPaxos with single relay layer. Static relays represent PigPaxos with non-random, predifined relay in each group.}
	\label{fig:aggregation_factor}
\end{figure}

The performance of PigPaxos is controlled mainly by the number of relay groups used for message dissemination and aggregation. A basic strategy for determining the number of relay groups in a scale-free manner is to use $\sqrt N$ as the number of groups. For example, $N=25$ implies using 5 groups with 5 nodes each. This makes the leader communicate with 5 relay nodes and the relay nodes to talk to 4 of their peers, a seemingly load-balanced setup. However, this naive strategy works well if the relays are static. PigPaxos, however, uses a uniformly random relay rotation and requires a different strategy for computing the best number of relay groups. Figure~\ref{fig:aggregation_factor} we show the maximum throughput of a 25 node PigPaxos cluster with and without the random relay node rotation. In this experiment, we used PigPaxos with partial response optimization configured to not wait for a single slowest follower in the relay group. Please note that the protocol without random relay rotation does not have the same liveness properties. 

We find that using $\sqrt N$ as a proxy for the number of groups works well when the relay nodes are static and do not change, as this configuration creates almost as much communication at relays as at the leader. However, the best throughput is achieved with random relay node rotation and the smallest number of relay groups possible. Such a setup minimizes the communication bottleneck at the leader the most and shifts a large portion of the message exchange overheads uniformly across the followers. The relay nodes do not get overwhelmed when talking to 23 other nodes in the relay group because the relays are rotated at every round. As relay nodes randomly rotate, the extra load in one round is offset by other concurrent rounds in which the node no longer serves as a relay. We discuss and analyze this observation further in Section~\ref{sec:num_relays_disc}.

%

\subsection{Latency and Throughput}
\label{sec:lat_tp}
Using relays adds a latency overhead. To study the impact the relay nodes have on PigPaxos latency, we compare PigPaxos latency with that of a corresponding traditional Paxos deployment. We also compare with Egalitarian Paxos (EPaxos)~\cite{epaxos}, since it presents an alternative way of reducing the leader bottleneck --eliminating dedicated leaders altogether. We run EPaxos with no conflicting operations to make sure we capture its absolute best performance. The benchmark clients are configured to communicate with Paxos and PigPaxos leaders for all operations, and with a random node in EPaxos on every operation.


\begin{figure}[t]
	\centering
	\includegraphics[width=\columnwidth]{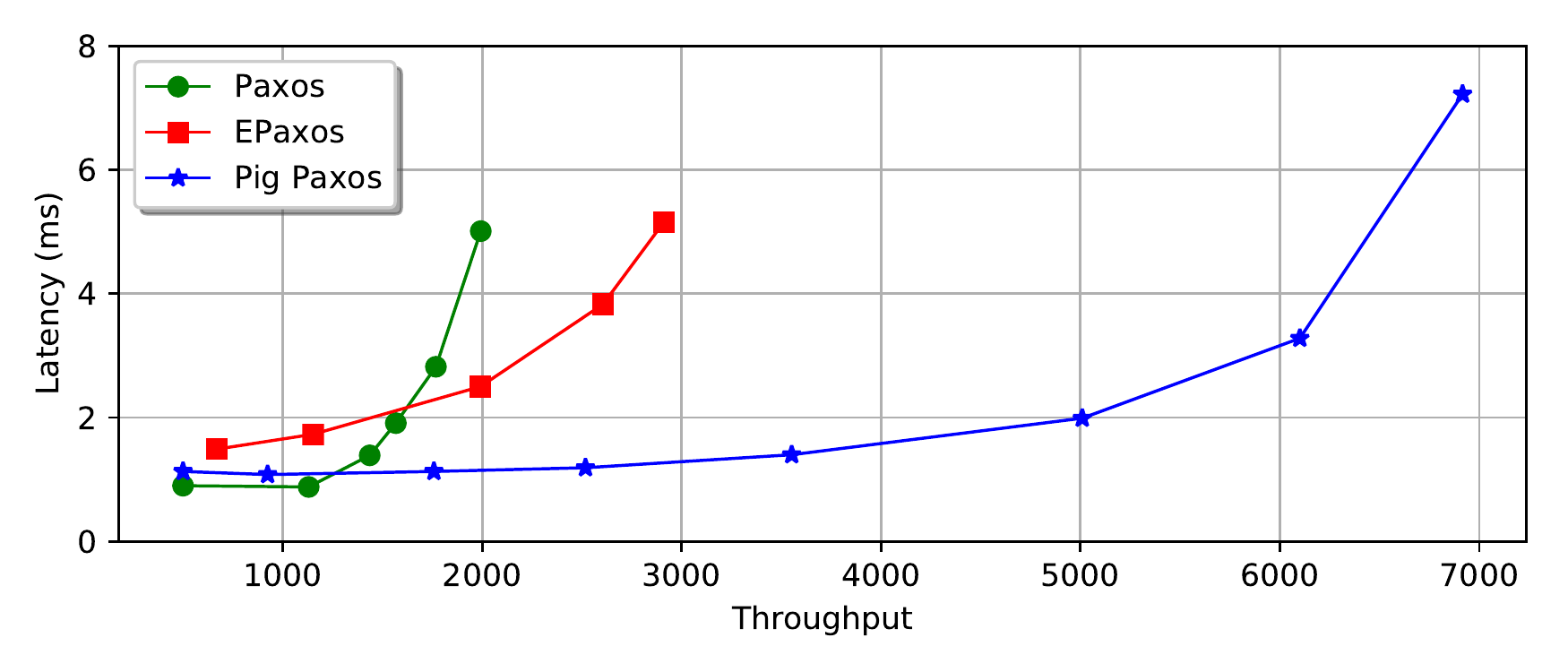}
	\caption{Latency and Throughput on 25-node cluster with 3 relay groups.}
	\label{fig:lat_tp_25}
\end{figure}

Figure~\ref{fig:lat_tp_25} illustrates the latency and throughput performance of these three protocols in a 25-node cluster. At this cluster size, EPaxos shows better scalability than Paxos because it allows all nodes to become leaders and serve operations. This removes the single leader bottleneck and allows EPaxos to have 50\% more throughput in conflict-free cases. On the other hand, EPaxos is significantly limited by its increased communication costs.
Since it needs to use larger quorums (fast quorum size = $3N/4$) before committing operations, it incurs more communication and higher latency to reach consensus.
The message size complexity of EPaxos also depends on the cluster size $N$, as the protocols need to keep track of dependencies coming from each node. This reliance on the cluster size for message handling greatly reduces the scalability of the entire cluster as $N$ grows. For instance, we have conducted a small experiment to see the overhead of more nodes in EPaxos and found that messages for a 25 nodes cluster take roughly 4 times more time to serialize than messages for a 5 nodes cluster.
%
%

Paxos, on the other hand, is limited by a single leader exchanging messages with all the followers. Initially, Paxos has a lower request latency, but its throughput gets saturated quickly and reaches its limit of around 2000 requests per second. While PigPaxos pays the initial price of having higher latency, we see that it scales to a much higher throughput with little latency deterioration after that. 

\begin{figure}[t]
	\centering
	\includegraphics[width=\columnwidth]{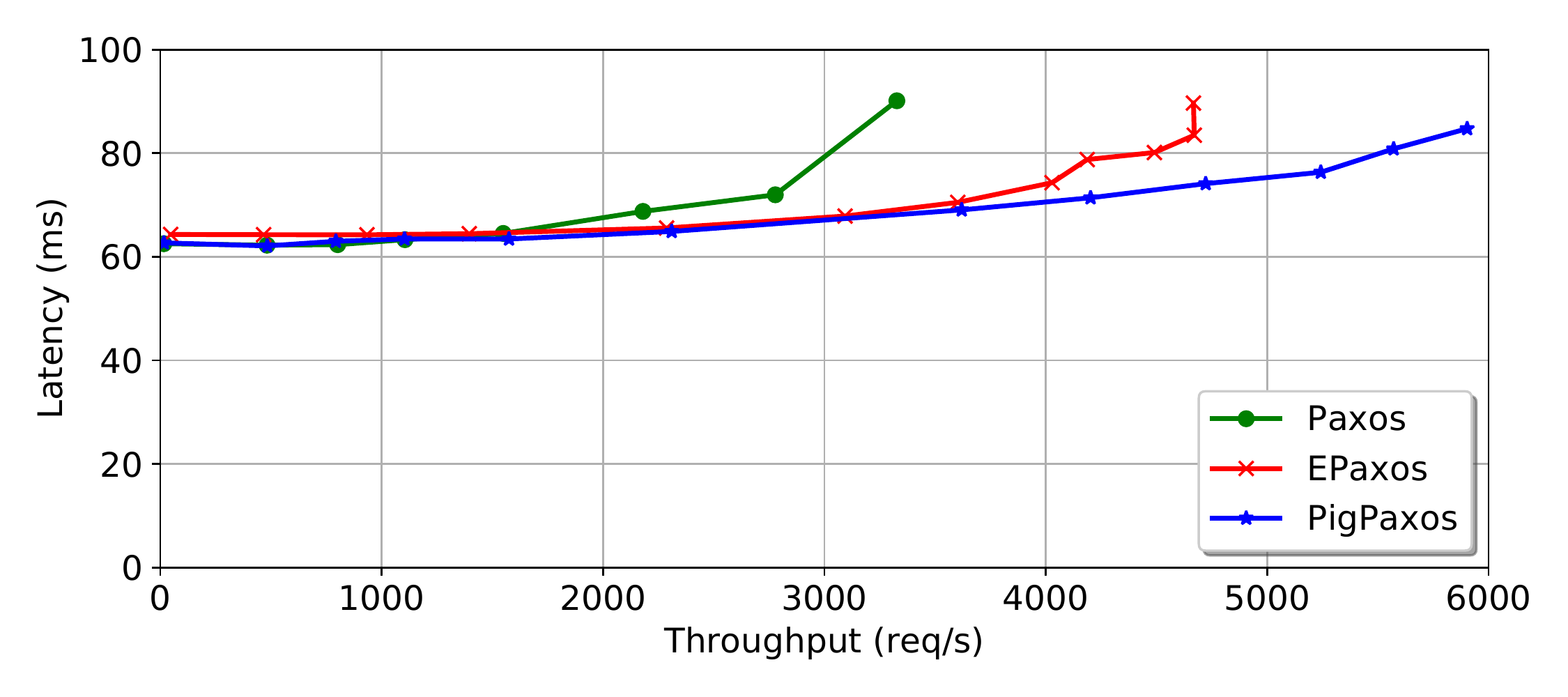}
	\caption{Latency and Throughput on 15-node WAN cluster in Virginia, California and Oregon.}
	\label{fig:lat_tp_15_wan}
\end{figure}

PigPaxos provides great throughput improvements over traditional Paxos in wide-area deployments as well. In Figure~\ref{fig:lat_tp_15_wan} we illustrate PigPaxos in 15 nodes configuration spread over AWS Virginia, California and Oregon regions. Each region represents a separate PigPaxos relay group, with the leader node located in the Virginia region. We deployed the clients in the same regions as the leader node. For EPaxos, we deployed clients in all regions and configured them to only communicate with nodes in their local regions. In WAN, the latency is dominated by cross-region distance, and the difference between Paxos and PigPaxos is not observable at low loads. EPaxos latency is slightly higher than both Paxos and PigPaxos due to a larger quorum and geographical properties -- depending on the region of request origin, the distance to reach quorum differs by approximately 5 ms in this setup. Similar to the local area experiments, PigPaxos maintains low latency for much higher levels of throughput.

PigPaxos also showed higher throughput than EPaxos in WAN, as EPaxos gets saturated quicker by a more complicated messaging and processing due to the dependency tracking, despite our workload having no conflicts. PigPaxos messages are also more complicated at the leader, however, smart aggregation allows us to keep the deserialization costs down compared to EPaxos that must include additional information in each message for dependency tracking. In a separate serialization microbenchmark written with Go's testing/benchmarking capability, we have estimated that PigPaxos' aggregated \textit{P2b} serializes about 14\% and deserializes about 8\% quicker than the similar \textit{PreAcceptReply} message of EPaxos. Other messages types have even bigger serialization performance differences, making every EPaxos node to spend more time in message handling than the PigPaxos leader.

\subsection{Small Clusters}
\begin{figure}[t]
	\centering
	\includegraphics[width=\columnwidth]{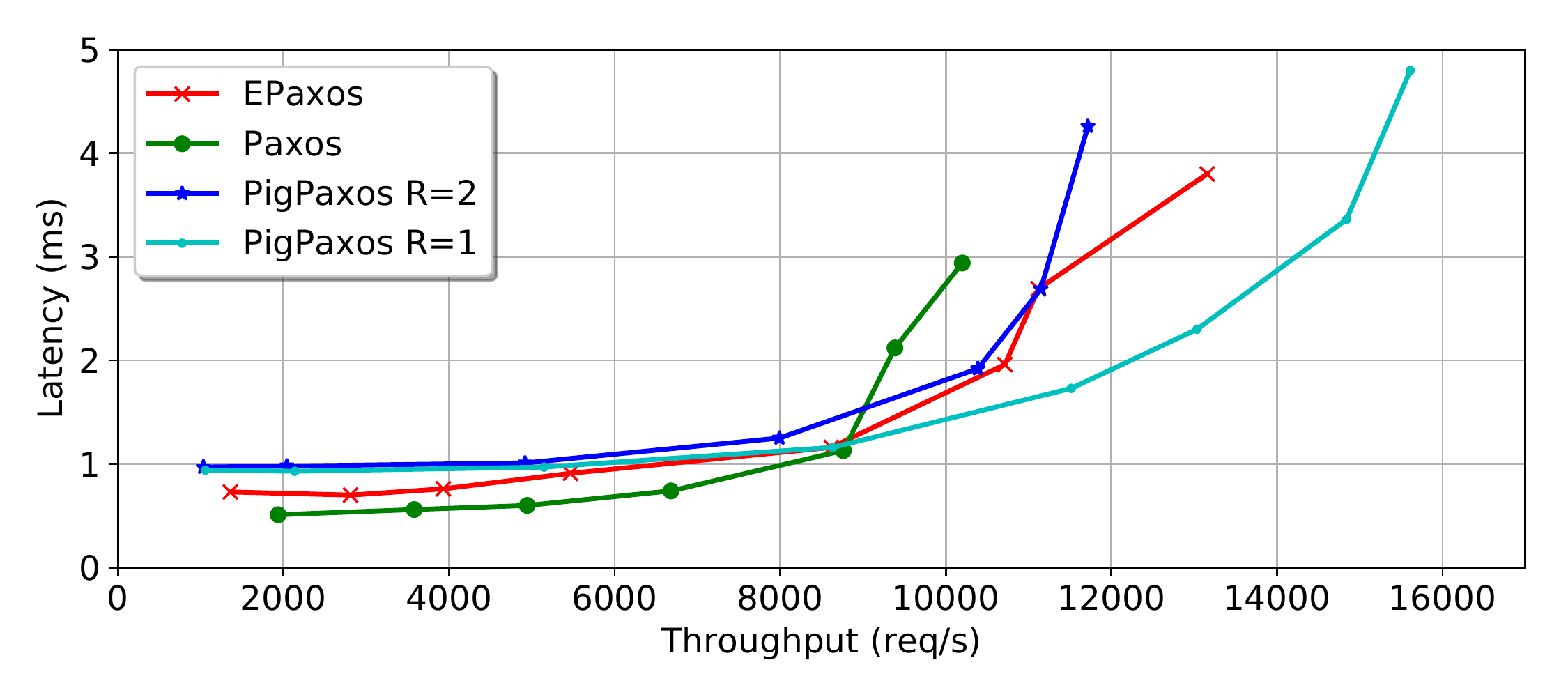}
	\caption{Latency and Throughput on 5-node cluster with 2 relay nodes.}
	\label{fig:lat_tp_5}
\end{figure}

We find that the benefits from PigPaxos extend to small clusters as well. In Figure~\ref{fig:lat_tp_5} we show a 5 nodes cluster running Paxos and PigPaxos with 1 and 2 relay groups ($R=1$ and $R=2$). For $R=1$ case, we used the single relay node optimization to wait for only the majority of nodes to aggregate at relay before retransmitting back to the leader. The difference in latency between the Paxos and PigPaxos protocols is more pronounced in this setup, as Paxos can maintain its low-latency performance for longer. However, even in this smaller cluster, PigPaxos scales to higher throughput than Paxos since the leader communicates with fever nodes.

In this small cluster, EPaxos shows its full potential, as it is not yet bottlenecked by expensive messages and can take advantage of its opportunistic leadership of operations. EPaxos shows around 12\% higher maximum throughput than PigPaxos for $N=5$ with $R=2$. A single relay group PigPaxos configuration ($R=1$) still outperforms all protocols, since it can spread the load in the cluster nearly evenly between all nodes without the penalty of conflict detection, resolution, and high message complexity.

\begin{figure}[t]
	\centering
	\includegraphics[width=\columnwidth]{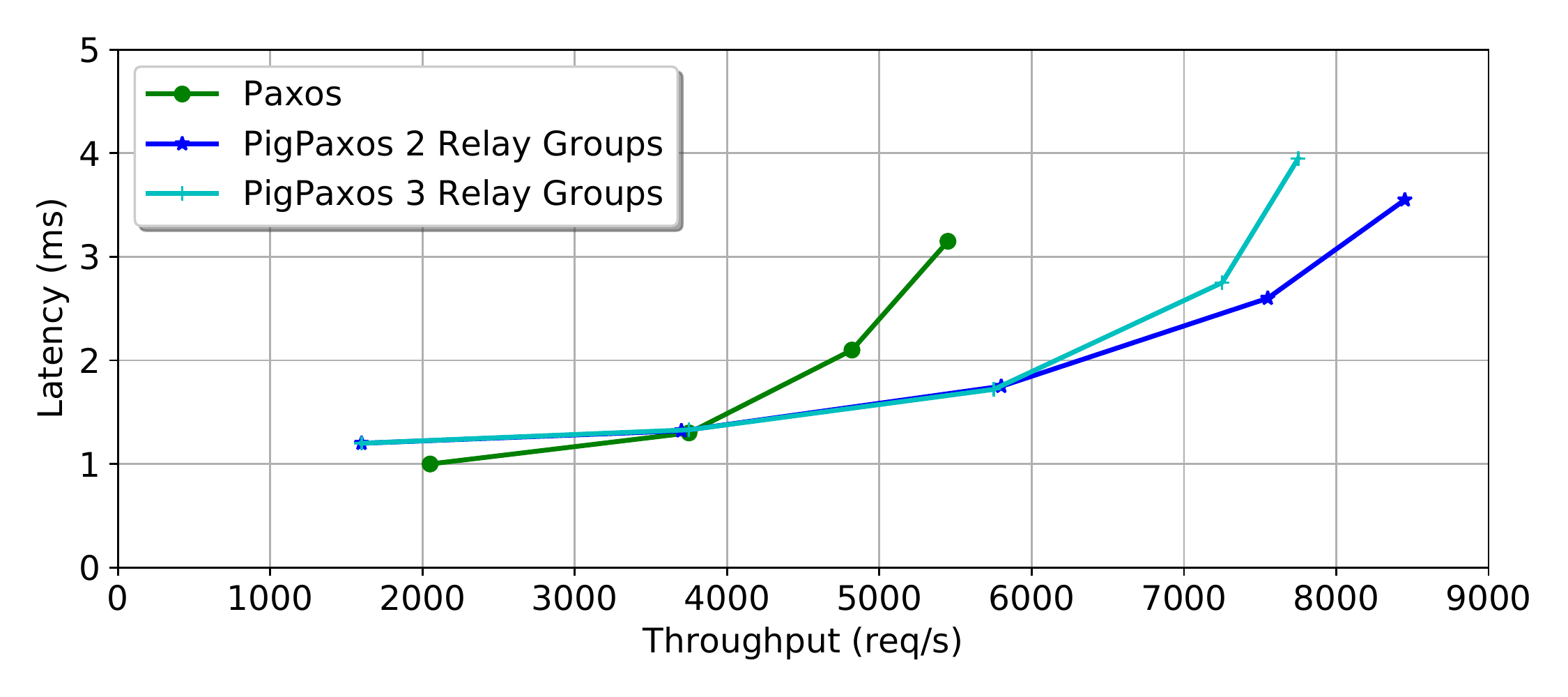}
	\caption{Latency and Throughput on 9-node cluster with 2 and 3 relay groups.}
	\label{fig:lat_tp_9}
\end{figure}

We repeat the experiment on Paxos and PigPaxos in a slightly larger 9 nodes cluster with $R=2$ and $R=3$ configurations, as shown in Figure~\ref{fig:lat_tp_9}. This setup is still relatively small but allows us to use higher values of $R$. As before, PigPaxos scales better than Paxos in both 2 and 3 relay groups deployments. However, with a bigger cluster size, Paxos's latency advantage over PigPaxos diminishes quickly even at lower throughput levels.

\subsection{Payload Size}
Payload size may have an impact on the communication performance of the system. Large messages require both more resources for serialization and more network capacity for transmission. To study how different payload size impacts the performance of PigPaxos and Paxos, we experiment with 25 node clusters, where PigPaxos uses 3 relay groups. To gauge the performance and scalability with respect to payload size we measured the maximum throughput on each system under a write-only workload generated by 150 clients running on 3 VMs. 

\begin{figure}[!t]
\centering
\begin{subfigure}{0.45\textwidth}
\includegraphics[width=\linewidth]{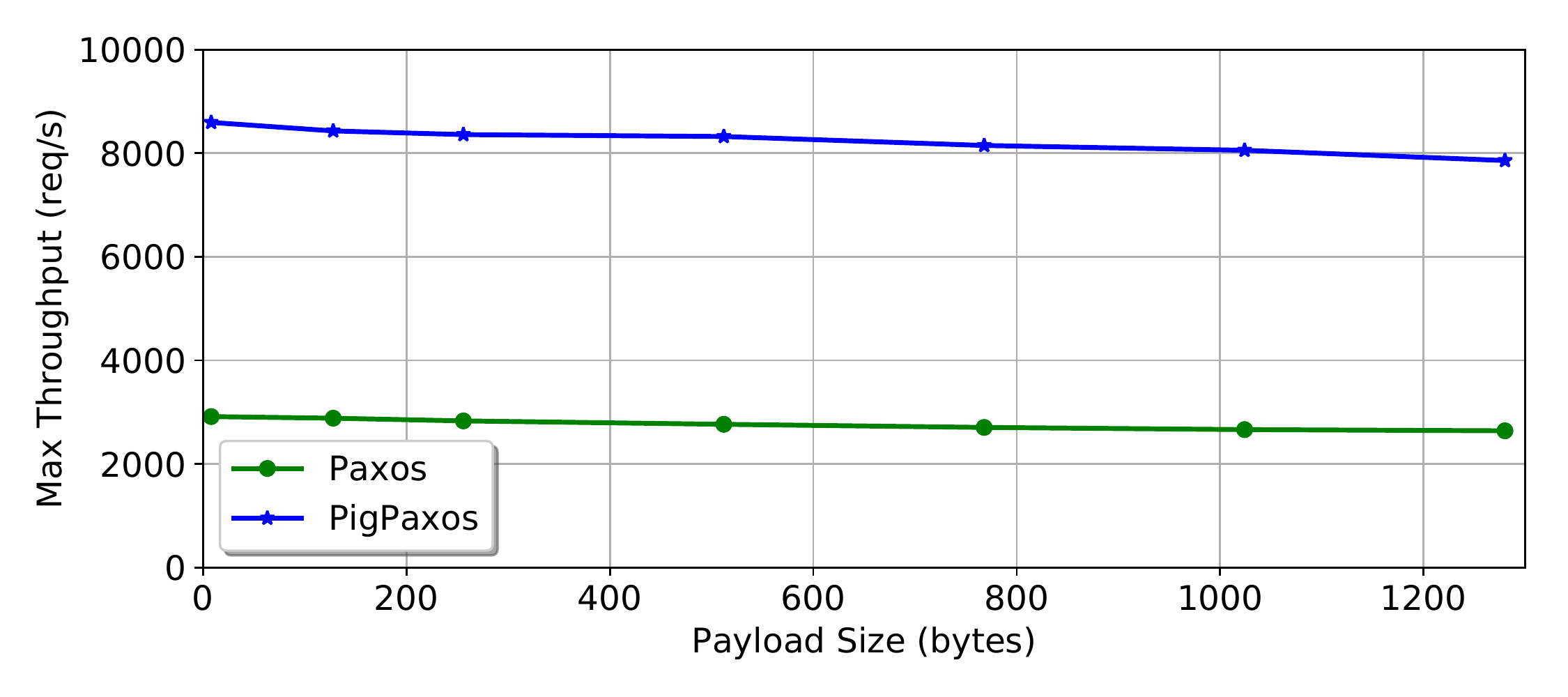}
\caption{Maximum throughput at various payload sizes.}
\label{fig:payload_tp_raw}
\end{subfigure}

\begin{subfigure}{0.45\textwidth}
\includegraphics[width=\linewidth]{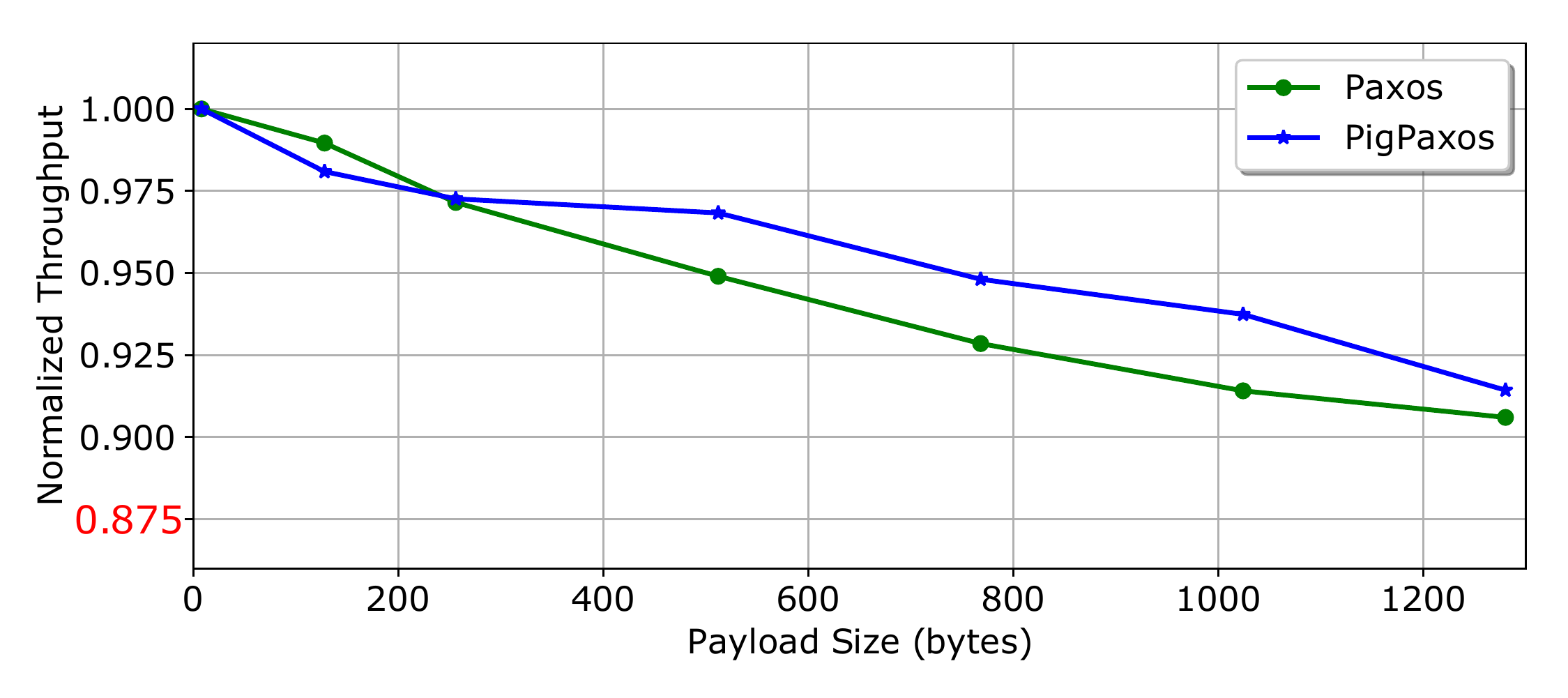}
\caption{Normalized throughput at various payload sizes. Note that the throughput scale starts at 0.86 of maximum throughput, and that neither protocol dipped below 0.9 of its top performance. }
\label{fig:payload_tp_norm}
\end{subfigure}
\caption{Performance of PigPaxos at various payload sizes, varying from 8 to 1280 bytes.}
\label{fig:payload_tp}
\end{figure}

Figure~\ref{fig:payload_tp_raw} shows the maximum throughput of PigPaxos and Paxos at payload sizes varying from 8 to 1280 bytes. While PigPaxos show three times more throughput than Paxos at all payload sizes, both protocols exhibit a similar relative level of degradation as the payload size increases. Figure~\ref{fig:payload_tp_norm} illustrates the throughput normalized to the maximum observed value.

\subsection{Partial Response Collecton in Steady-State}

\begin{figure}[t]
	\centering
	\includegraphics[width=\columnwidth]{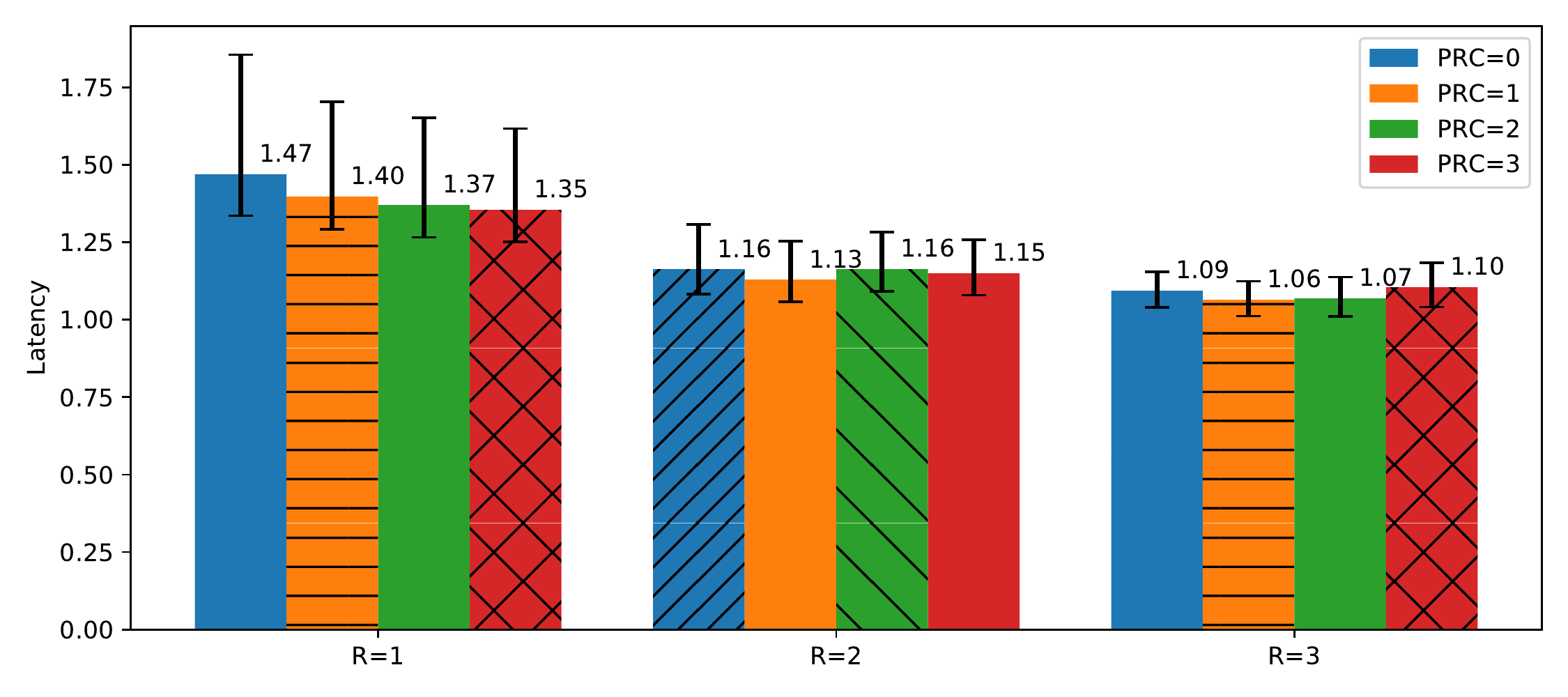}
	\caption{Median and IQR latency with different configurations of partial response collection with different numbers of relay groups in a 25 nodes cluster. $PRC=x$ indicates that the relay is not waiting on $x$ slowest replicas to send an aggregated response to the leader.}
	\label{fig:pig_prc_nofailure}
\end{figure}

Partial response collection optimization allows a relay group to respond to the leader without waiting on one or more of the slowest nodes, slightly improving the latency in the steady operation and allowing to better tolerate occasional straggler nodes. In Figure~\ref{fig:pig_prc_nofailure} we illustrate the impact of this optimization on median and IQR (25th and 75th percentile) latency. For this experiment, we ran the benchmark at a fixed workload of 3000 requests/second over a 300 seconds interval. We see the real benefit from this feature in the steady-state in a configuration with just a single relay group. In such a setup, a relay needs to wait on 23 other nodes before replying, making an occasional straggler slow down the system In the configurations with multiple relay groups, it is sufficient to have an entire relay group to not reply and the leader still reaching the majority, making the benefit of masking a slowdown in a few occasional nodes not needed when a slowdown of an enitre relay group can be tolerated.

\subsection{Performance Under Failures}
Since PigPaxos preserves the control and decision-making logic in Paxos,
it can make progress as long as the majority of nodes are still up and running. However, it is also important to maintain an adequate level of performance even under failures, and with the use of relay groups and relay nodes, it is worth investigating whether this property is satisfied as well.

\begin{figure}[t]
	\centering
	\includegraphics[width=\columnwidth]{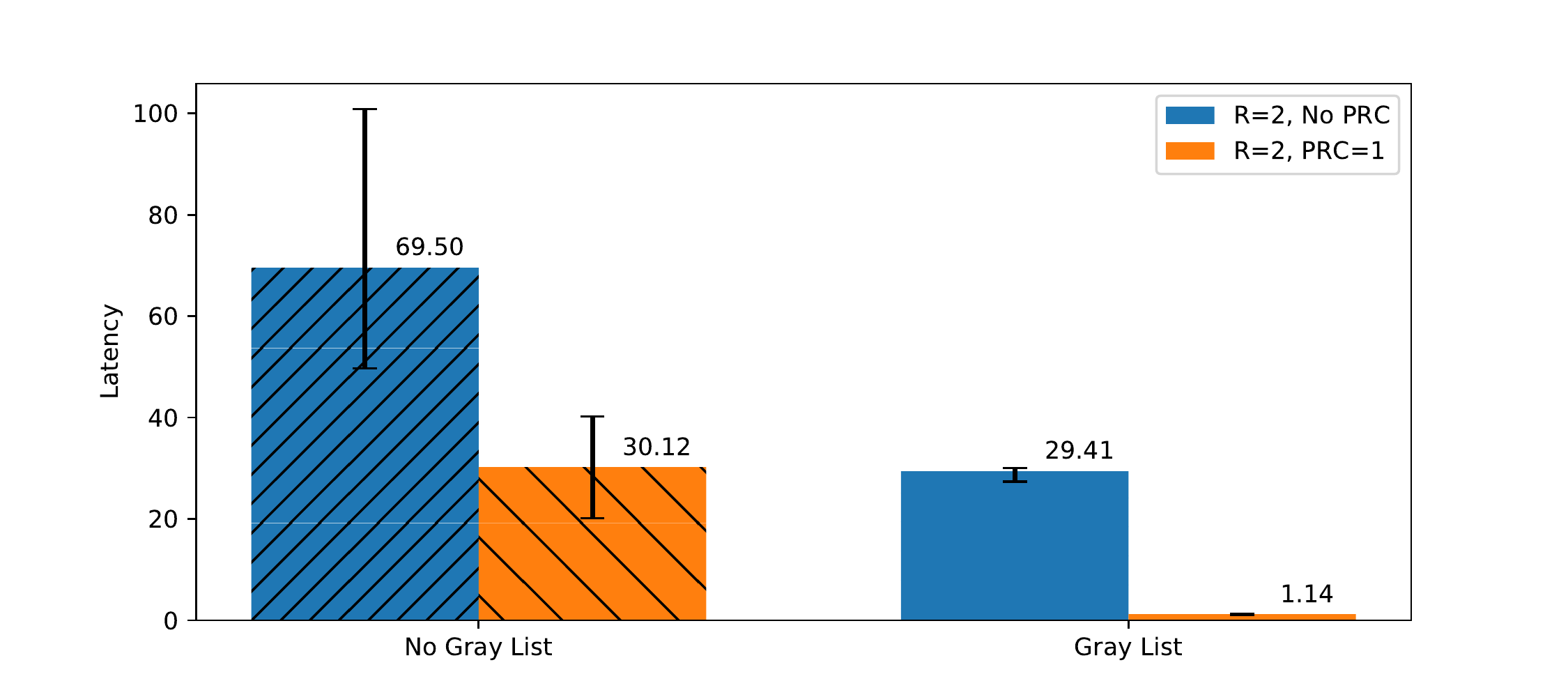}
	\caption{Median and IQR latency using gray lists at different levels of partial response collection in a 25 nodes cluster with 3 relays groups($R=3$) and one node failure. $PRC=x$ indicates that the relay is not waiting on $x$ slowest replicas to send an aggregated response to the leader.}
	\label{fig:pig_prc_failure}
\end{figure}

Partial response collection and gray lists allow PigPaxos to maintain reasonably high performance while tolerating some faults. A fault in PigPaxos prevents a relay group from collecting all the responses, causing a timeout in the relay group, unless we use partial response collection. If other relay groups are sufficient to form the majority quorum, the system will proceed as normal, however, if the faulty relay group is required for the majority, then every operation will incur a timeout, as the relay is waiting on the faulty node. Moreover, some operations will incur a higher timeout when a faulty node is chosen as the relay, causing the timeout and retry at the leader. This is evidenced in Figure~\ref{fig:pig_prc_failure} in configuration with $R=2$ and no partial response collection without the gray lists.  In this figure, we show median latency from a steady workload generated by 30 clients in a 25 nodes cluster with one node failure. 

Adding partial response collection to the system allows us to avoid the timeout due to waiting on the faulty node at the relay, but does not prevent cases when the leader picks the faulty replica as the relay. Such configuration in Figure~\ref{fig:pig_prc_failure} ($R=2, PRC=1$, no gray lists) shows much better latency than no partial response collection. However, the latency is still significantly higher than the no-fault operation. This is because one operation that goes through a timeout stalls the entire pipeline or log, and causes all higher-numbered operation to delay as well to preserve linearizability. 

However, we can also reduce the impact of picking a faulty node as a relay by keeping track of nodes suspected to be down with our gray list (Section ~\ref{sec:gray}) optimization. By excluding problematic nodes from being frequently selected as relays, we avoid such timeouts and pipeline stalls, as evidenced in Figure~\ref{fig:pig_prc_failure} gray list configurations. When combining partial response collection and gray lists we can achieve median latency comparable to the fault-free case.

\begin{figure}[t]
	\centering
	\includegraphics[width=\columnwidth]{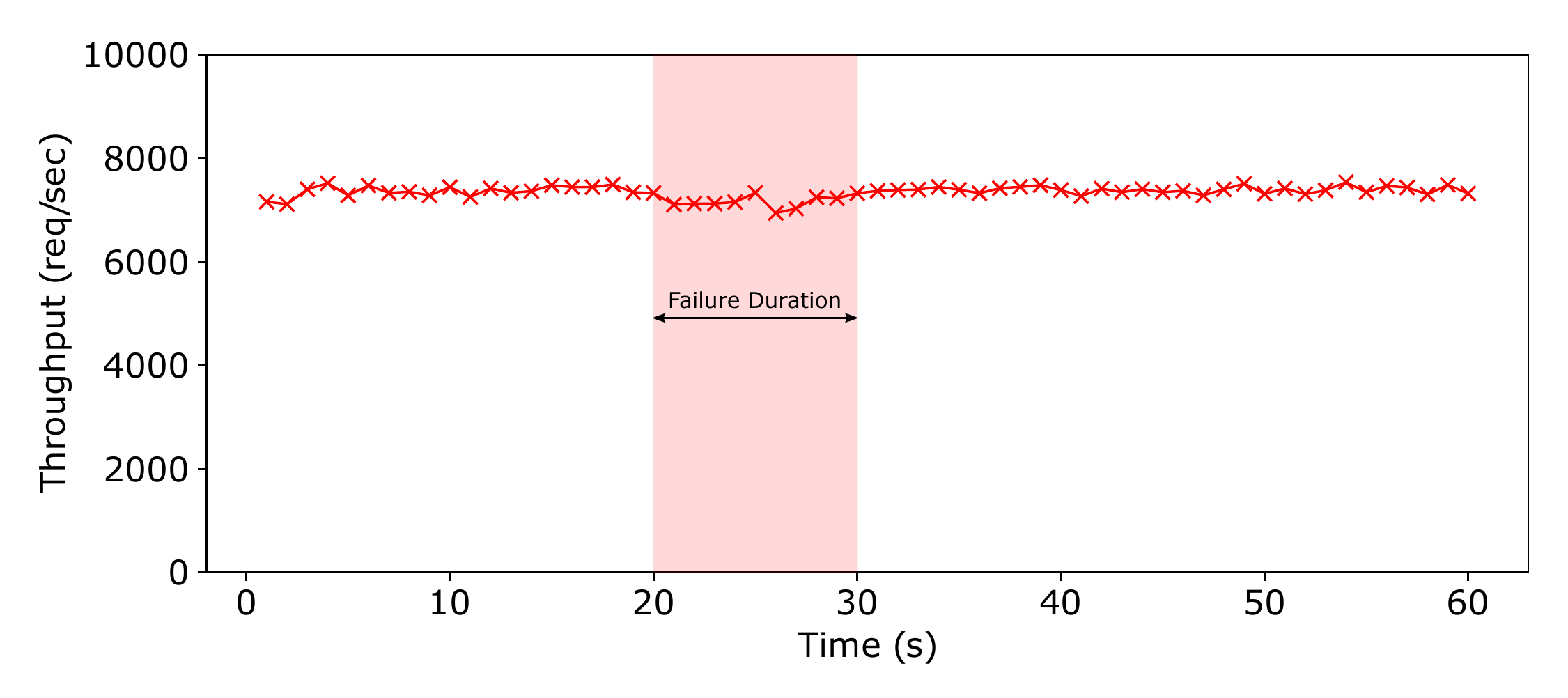}
	\caption{Maximum throughput on 25 nodes cluster with 3 relay groups under a single group failure. Throughput is sampled over 1-second intervals.}
	\label{fig:3aggregators1failure}
\end{figure}

Figure~\ref{fig:3aggregators1failure} illustrates how the maximum throughput of PigPaxos is impacted when one of the 3 relay groups in a 25 node cluster is faulty with multiple nodes not responding. In this experiment, we did not use any additional optimizations beyond the group timeout to illustrate the worst possible case of a fault relay group. The timeout was set to 50ms, more than 40 times over the normal operation latency.

A failure in a single relay group causes the relay group to timeout. However, the two other relay groups still constitute the majority of nodes and can respond to the leader quickly, resulting in very little change in performance. We find that the maximum throughput over the duration of the fault declined by only ~3\%, mainly due to the leader having to wait a little longer.


\section{Discussion}
\label{sec:disc}
\subsection{The Leader Bottleneck and the Number of Relay Groups}
\label{sec:num_relays_disc}
PigPaxos relies on the relay groups to deliver messages to every participating node. As we have observed in Figure~\ref{fig:aggregation_factor}, the number of relay groups in a deployment impacts the performance significantly. For example, with two relay groups ($R=2$), PigPaxos achieved ~11\% higher throughput than the $R=3$ configuration. To explain this result, we refer to the performance modeling ideas in Ailijiang et al.~\cite{dissecting_perf} and adapt and extend that modeling idea to cover PigPaxos. In particular, we adapt the model to look at the number of exchanged messages $M$ in every node of the PigPaxos protocol as a proxy for the load on the node. A leader needs to receive one message from the client and eventually send a reply, besides, the leader communicates with $R$ number of relay groups. The total number of messages $M_l$ handled by the leader is then expressed as:
\begin{align}
M_l = 2R + 2 \label{eqt:ml}
\end{align}

The message load on the followers depends on the role it performs for a given round: relay followers are responsible for vastly more messages than the regular follower nodes. A non-relay processes only one round trip communication with the relay node.
A relay node, on the other hand, needs to receive one incoming message from the leader, send one message back to the leader, and handle a round trip communication with every remaining follower in the relay group. However, the random alternation of the relay nodes shields them from becoming hotspots: the extra traffic load a relay node incurs in one round is offset in consecutively-numbered concurrent rounds when the node no longer serves as a relay. Therefore, on average, we have the following message load on each follower:
\begin{align}
M_f = 2\left(\frac{R}{N-1}\right)\left(\frac{N-R-1}{R}\right) + 2 \\
    = \frac{2\left(N-R-1\right)}{N-1} + 2 \label{eqt:mf}
\end{align}

\noindent
where $N$ is the total number of nodes in the system, and $\frac{R}{N-1}$ is the probability of a node being chosen as a relay for the round. 

This analysis of the number of message exchanges at the nodes paints a similar picture to our empirical results, indicating that to eliminate bottleneck for the system it is best to keep the number of relay groups, $R$, small.
The reason for this is that the leader still handles more communication and data processing than the ``average'' follower node. For example, in Table~\ref{tab:msg_overhead} we show the number of messages handled by the leader and followers for a different number of relay groups in a 25 nodes cluster. We also compute the ratio of messages processed by the leader to that of the followers. We observe a dramatic increase in this ratio as the number of relay groups increases, indicating that the load on the leader and the followers becomes more disproportionate with additional relay groups. Such disbalance reaches a factor of 25 times for a regular Multi-Paxos deployment. We have confirmed this growing disparity between load at the leader and followers by measuring the CPU utilization and found a widening difference in CPU usage between the leader and the followers as the number of relay groups increases. This matches with the empirical results in Section~\ref{sec:num_relays}.

\begin{table}[]
\resizebox{\columnwidth}{!}{
\begin{tabular}{|l|l|l|l|}
\hline
\# of Relay Groups ($R$) & Messages at Leader ($M_l$) & Messages at Follower ($M_f$) & $M_l/M_f$ \\ \hline
1 & 4&  3.92 &  1.020\\ \hline
2 & 6&  3.83 & 1.567\\ \hline
3 & 8&  3.75 & 2.132\\ \hline
4 & 10&  3.67 & 2.725\\ \hline
5 & 12&  3.58 &  3.352\\ \hline
6 & 14&  3.50 &  4.0\\ \hline
\textbf{24 (Paxos)} & \textbf{50} & \textbf{2} &  \textbf{25.0} \\ \hline
\end{tabular}
}
\caption{Message load at leader and followers for different number of relay groups in 25 nodes cluster. }
\label{tab:msg_overhead}
\vspace{-2mm}
\end{table}

\begin{figure}[t]
	\centering
	\includegraphics[width=\columnwidth]{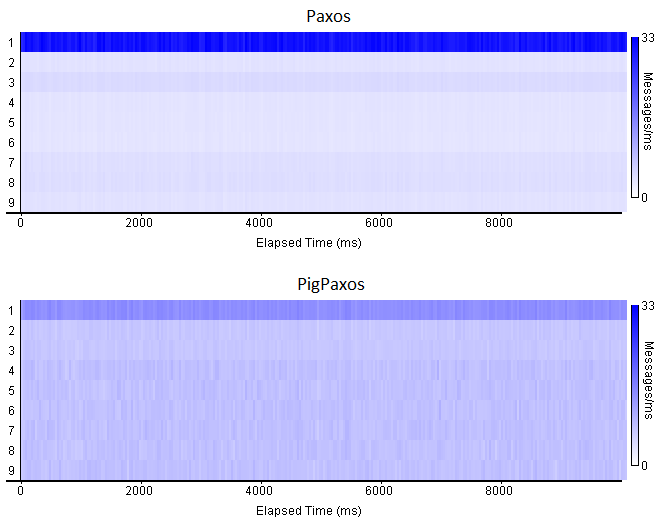}
	\caption{Heatmap of in-flight messages between nodes in 9 nodes Paxos and PigPaxos cluster on AWS EC2 under 1500 requests per second workload. Node 1 is a leader in both cases. PigPaxos used 3 relay groups. Every in-flight message is counted upon the first observation in Retroscope.}
	\label{fig:retro_msg_inflight}
	\vspace{-2mm}
\end{figure}

We have also verified the message load experimentally using Retroscope~\cite{retroscope} to collect, among other things, the in-flight message information. Figure~\ref{fig:retro_msg_inflight} illustrates a heatmap of in-flight messages to and from each node in a 9-server AWS cluster under a 1500 req/s workload. In the figure, the color depth represents the communication intensity, with the maximum message exchange rate clocked at 33 messages in one millisecond. We can clearly observe that PigPaxos distributes the communication load in the cluster better compared to Multi-Paxos, even though the leader still remains a communication bottleneck. 

\subsection{Single Relay Group Configuration}
As evidenced from our evaluation (Figures~\ref{fig:aggregation_factor} \&~\ref{fig:lat_tp_5}), single relay groups perform admirably well in PigPaxos, and one may ask a question of whether any other configuration is required at all. The major drawback of $R=1$ PigPaxos is the potential for slightly higher tail latency under failures. When the leader picks a crashed or slow node as a relay, it will stall the request for the timeout duration, which will translate into the tail latency. At the same time, $R>1$ configurations often reach the majority with one relay group not replying at all, seamlessly masking such failures. Under more severe failure conditions that affect multiple relay groups, there will be extra latency due to the leader having to wait for the timeouts at the relays.

Another area where $R>1$ PigPaxos shines is geo-distributed replicated state machines. Having each datacenter or region to be a separate relay group means that the payload is replicated to each region only once, conserving the cross-region bandwidth and reducing costs.
This efficiency comes from the ability to assign all nodes in the region to a single relay group and making PigPaxos leader send only one message per region across the wide-area network (WAN). With many cloud providers charging for WAN traffic, a PigPaxos system can provide significant savings to applications requiring geo-redundancy and strong consistency. For example, a 3 region deployment with 3 nodes in each region will have only 2 messages going across WAN for each write operation. A paxos-backed system, in contrast, will send 6 separate messages, representing 3 times the WAN traffic. In a geo-distributed environment, this $R$ equals the number of regions setup also means that the majority quorum is reached by nodes in the datacenters closest to the leader, while $R=1$ setup may rotate the relay node to a far away datacenter causing higher latency.

\subsection{PigPaxos in Small Clusters}
PigPaxos scalability benefits apply not only to large clusters but also to smaller deployments. As we illustrate in Figure~\ref{fig:lat_tp_9}, a 9 node PigPaxos cluster with $R=2$ improves the throughput over classical Multi-Paxos by as much as 57\% with very little degradation in latency, and even more, performance is possible with $R=1$. Moreover, the benefit of PigPaxos extends down to 7 and even 5 node clusters (Figure~\ref{fig:lat_tp_5}). In the latter case, PigPaxos has very little flexibility for its configuration, as the only options are to use one or two relay groups. However, in both situations, we have seen performance improvements compared to the regular Multi-Paxos. Using the analytical approach described earlier, Table~\ref{tab:msg_overhead_small} illustrates the relative message-induced bottleneck of Paxos and PigPaxos leaders in a small 5 nodes cluster.

\begin{table}[]
\resizebox{\columnwidth}{!}{
\begin{tabular}{|l|l|l|l|}
\hline
\# of Relay Groups ($R$) & Messages at Leader ($M-l$) & Messages at Follower ($M_f$) & $M_l/M_f$ \\ \hline
1 & 4&  3.5 & 1.142\\ \hline
2 & 6&  3 & 2.0\\ \hline
4 (Paxos) & 10 & 2 & 5.0 \\ \hline
\end{tabular}
}
\caption{Message load at leader and followers for different number of relay groups in a small 5 nodes cluster.}
\label{tab:msg_overhead_small}
\vspace{-2mm}
\end{table}


\subsection{PigPaxos Overheads}
Pig protocol aims to shift the load in fan-out/fan-in concurrent communication away from a single node that initiates communication. PigPaxos adopts this pattern to reduce the leader bottlenecks and make the message handling/processing costs more even in the entire cluster. However, in the process, Pig introduces several overheads and performance compromises. Most importantly, Pig splits a direct communication between the leader and the followers into a two-hop message exchange. This inevitably leads to a latency increase. For example, in Figures~\ref{fig:lat_tp_25}, \ref{fig:lat_tp_9}, and \ref{fig:lat_tp_5} we can observe Multi-Paxos having as much as ~40\% lower latency when the workload's throughput is low. 

Another important consideration is the message size in Pig. In our experimentation, we have established that bigger messages take longer to serialize and deserialize than smaller messages. While the leader in PigPaxos processes fewer messages, the aggregated responses are naturally bigger than the individual responses in Multi-Paxos, erasing some of the benefits of Pig. However, in PigPaxos we make sure that the aggregated message is as small as possible by removing all redundant information. For instance, for Paxos leader to recognize double message delivery and prevent double-counting, it needs to keep track of all the nodes that have voted. However, most often the nodes will vote one time, so when replying to the leader, the relay node performs initial deduplication and only includes the IDs of followers who did not vote. This list of missing nodes is much smaller than the list of all followers who voted, but it is sufficient for the leader to prevent double-voting.

Finally, as Pig changes the communication flow in the cluster, there is a possibility to create an overall different number of messages flowing in the network compared to the traditional Multi-Paxos. More messages per communication round may mean a decreased efficiency, despite better load balancing. Luckily, it is trivial to compute the total number of messages sent in each communication round of PigPaxos: a leader sends $R$ messages to the relays and $1$ message as a response to the client, each relay sends $\frac{N-R-1}{R}$ messages to the followers and $1$ message as an aggregated response to the leader. Followers send just $1$ message each. Adding this all together and simplifying yields: $2N-1$ messages, a number that does not depend on the relay groups and remains constant for a fixed size cluster. As such, Pig allows us to achieve a better balance between message handling at different components without an increase in the total number of messages required.

\subsection{Number of Relay Layers}
\label{sec:number_relay_layers}
While we limited our discussion to a single layer of relay nodes, in the most generic form, Pig communication can be extended to multiple layers of relay nodes. Multi-layered Pig pattern may allow lower layer relays to offload some of the communication burdens from the upper layers. However, this only makes sense if there is no bottleneck on the upper levels of the communication flow. As we have shown for the 25 nodes cluster, the leader remains the bottleneck even with just a single relay node.



We further find that even in the most extreme configurations with $R=1$ and an arbitrary number of follower nodes, the leader remains a bottleneck. The leader's message load $M_l$ is a linear function of the number of relay groups $R$, as shown in formula~\ref{eqt:ml}. To minimize the leader message load, we need to minimize the number of relay groups, which cannot be smaller than 1. As a result, the smallest message load on the leader is $M_l = 4$. At the same time, with $R=1$ and $N\to\infty$, the load on the followers $\frac{2(N-R-1)}{N-1} + 2$ asymptotically increases to 4, as $\lim_{N \to \infty} \frac{2(N-2)}{N-1} + 2 = 4$.
In other words, even in the most extreme case, the average/amortized load on follower nodes only matches that of the leader, indicating that the leader remains a bottleneck regardless of the cluster size and the number of relay groups.
Since the bottleneck cannot be shifted entirely to the followers with any $N$, adding more layers to offload the communication work from the followers will not result in a performance gain.

These being said, we want to mention a caveat: our simple model does not consider other variables that may play a significant role for very big clusters or very big messages. For example, in a large cluster, the cost of maintaining many connections may start to add up and affect the relays, leaving the possibility for multi-layer communication trees to be beneficial at a very large scale. 

\section{Related Work}
\label{sec:related}

\subsection{Flexible Paxos}  
Flexible quorums Paxos, or FPaxos~\cite{fpaxos}, observes that the majority quorum is not necessary for phase-1 and phase-2. Instead, it shows that the Paxos properties hold as long as all quorums in phase-1 intersect with all quorums in phase-2. This result enables deploying Multi-Paxos protocols with a smaller phase-2 quorum, providing better performance at the cost of reduced fault tolerance.

While using a smaller Q2 is beneficial for reducing latency, it does not reduce the leader bottleneck and does not help improving scalability or throughput. This is because flexible Paxos talks to all nodes and takes the first $|Q2|$ responses for satisfying phase-2 requirement, although the other responses also arrive and overwhelm the leader. While it is possible to use a thrifty optimization~\cite{epaxos} and contact only $|Q2|$ nodes for phase-2, in that case a single faulty or sluggish node in Q2 stalls the performance.

In contrast to the flexible quorums approach, PigPaxos clears the bottleneck at the leader while also preserving the robustness against failures. However, it is possible to employ the PigPaxos approach with the flexible quorums idea, for providing improved scalability and throughput to FPaxos~\cite{fpaxos}.



\subsection{Multi-leader consensus protocols}
Scalability bottlenecks have been a big problem in consensus protocols and have received a lot of attention. Mencius~\cite{mencius} uses rotating leaders to alleviate the bottleneck at a single leader to improve throughput. 
For Mencius to work effectively, the workload has to be partitioned across each leader nicely, so that each leader has something to propose in their turn. Stalls in rotating leader proposals create problems with Mencius performance, and delays and failures cause tricky corner cases.
In contrast, PigPaxos is simple to implement and it avoids these issues.
Instead of rotating leaders, PigPaxos rotates the small number of relay nodes the leader interacts with. This clears the communication bottlenecks and improves scalability and throughput, while still keeping the protocol simple and avoiding the complexity and performance complications Mencius suffers from.
%

Canopus~\cite{canopus} is a multi-leader consensus protocol aimed at wide-area deployments. It trades off latency and fault-tolerance for throughput. The protocol heavily relies on batching and requires a lot of communication steps to finish a single consensus round. Canopus divides the nodes into a fixed number of groups, called super-leaves. At the super-leaf level, Canopus assumes synchronized rounds and a network-based reliable broadcast primitive. On the first communication cycle, each node within the super-leaf exchanges the list of proposed commands with other super-leaf peers using the reliable broadcast primitive. Every node then orders these commands in the same deterministic order, creating a virtual node that combines the information of the entire super-leaf. In consecutive cycles, the virtual nodes exchange their commands with each other at increasingly higher levels until a single super-node emerges as the root of the tree. At that point, every physical node has all commands in the same order and consensus has been reached across all super-leaves. Canopus suffers from fault-tolerance problems, as it cannot tolerate a network partition or failure of any single super-leaf. Since the nodes that constitute a super-leaf need to be located within the same rack (for using the reliable broadcast primitive), Canopus is especially prone to super-leaf failures and loss of availability.
In contrast to Canopus, PigPaxos can tolerate failures of up to half of the nodes in the system. Moreover, PigPaxos achieves high-throughput without an observable increase in latency in most cases, as we show in Section~\ref{sec:eval}.



WPaxos~\cite{wpaxos} presents a WAN-optimized multi-leader Paxos protocol. WPaxos selects leaders (and maintains logs) to be per-object, and employs an object stealing protocol, with adaptive stealing improvements to match the workload access locality~\cite{migration_policies}. More specifically, multiple concurrent leaders coinciding in different zones steal ownership of objects from each other using phase-1 of Paxos, and then use phase-2 to commit update-requests on these objects locally until they are stolen by other leaders. To achieve fast phase-2 commits, WPaxos leverages the flexible quorums result~\cite{fpaxos} and appoints phase-2 quorum Q2 to be close to their respective leaders.
The multi-leader, multi-quorum setup in WPaxos helps with both WAN latency and throughput due to the smaller and geographically localized quorums.
In contrast to WPaxos that uses multiple leaders and multiple conflict domains, PigPaxos improves throughput and scalability using a single leader within a single conflict domain. However, PigPaxos optimization is still applicable within the framework of WPaxos.
In addition to their Q2 used for committing, WPaxos leaders can employ PigPaxos for implementing full replication to a large number of nodes and learning when these additional nodes commit as well. This technique could be useful for implementing bounded-staleness~\cite{manyfacesconsistency} at large-scale geo-replicated database deployments.

Fast Paxos~\cite{fastPaxos} and EPaxos~\cite{epaxos} removes the requirement of partitioning of conflict domains between nodes prescribed in the previous multi-leader approaches, and instead try to opportunistically commit commands.
%
%
Any node in EPaxos becomes an opportunistic leader for a command and tries to commit it by running a phase-2 of Paxos in a super-majority quorum system. If some other node in the super-majority quorum is also working on a conflicting command, then the protocol requires performing a second phase to record the acquired dependencies, and agreement from a majority of the Paxos acceptors to establish order on the conflicting commands  is needed.
EPaxos suffers from increased number of conflicting commands for heavy workloads, and its performance plummets.
In Section~\ref{sec:eval}, we give performance comparison of EPaxos and PigPaxos. 
In contrast to EPaxos, PigPaxos provides linearizability of all operations and clears communication bottlenecks and achieves high throughput within a more simple single leader Paxos implementation. 

Compartmentalized Paxos~\cite{compartmentalizedpaxos}, like a classical single instance Paxos~\cite{paxos}, breaks down Paxos protocol into multiple distinct roles and allocates these roles to separate nodes. The protocol introduces proxy-leaders, who are responsible for tallying votes and acquiring the majority of votes on the proposers' behalf. The stable proposer itself is only responsible for establishing the order of operations and selecting a different proxy-leader for each command. Proxy-leaders, in a similar manner, talk to one of many acceptor groups to commit the operation. A proxy-leader and an acceptor group can be generalized as a single relay group in PigPaxos. However, Compartmentalized Paxos requires learner nodes to maintain a single state machine, since no single proxy-leader or acceptor group knows of all operations. Similar to PigPaxos, the compartmentalized idea introduces additional failure modes,  and recovers by retries. For instance, the failure of a proxy leader will stall the consensus round and affect the performance of the protocol while a retry on another proxy leader is taking place. However, some failures, such as a crash of an entire acceptor group stalls the Compartmentalized Paxos indefinitely, as part of the log is lost.

%
%

\subsection{Weak consistent replication}

The 2005 paper by Saito and Shapiro~\cite{optimisticreplication} provides a comprehensive survey of the weak consistent replication protocols.
Weak consistent replication protocols promise higher availability and performance, but they allow replicas to temporarily diverge and let users see inconsistent data. In these techniques, updates are propagated in the background, and occasional conflicts are fixed after they happen. Since they require little synchronization among replicas, they are able to scale to a large number of replicas.

Weak consistent replication approaches adopt techniques such as version vectors, vector clocks, modified bits, Thomas's write rule, and are able to provide only bounded-divergence or eventual consistency guarantees, rather than strong consistency guarantees.
Weak consistency replication protocols use multiple concurrent leaders (by trading off consistency) to circumvent the single leader bottlenecks.
For communication, they use flooding, rumor mongering, and directed gossiping.
While the Pig communication primitive could also provide some performance improvements to weak consistency replication protocols, those would not be as pronounced.

\subsection{Blockchain protocols}
Blockchain protocols solve consensus in the presence of Byzantine nodes. Recent work on blockchain consensus protocols employ a tiered hybrid method, and choose a council among all participants~\cite{byzcoin_16,algorand_17} to reduce the size of consensus quorums, and then run Byzantine fault-tolerant protocols to achieve quick finality in a deterministic manner in this council.
In particular, the Practical Byzantine Fault Tolerance (PBFT)~\cite{pbft} protocol forms a basis/template for most Byzantine fault-tolerant (BFT) consensus protocols, such as Casper~\cite{casper} used in Ethereum, Tendermint~\cite{tendermint} used in Cosmos, and HotStuff~\cite{hotstuff} used in Facebook's Libra.
Since these protocols have similar protocol architecture to Paxos, the PigPaxos aggregation technique we introduced here can also be adopted to help improve scalability for these protocols. 

Several blockchain protocols~\cite{scal_bc_18,scal_brb_19} use gossip-based message dispersal instead of point-to-point communication.
FastBFT~\cite{scal_bc_18} introduces a novel message aggregation technique that combines hardware-based trusted execution environments with lightweight secret sharing, and reduces the BFT consensus message complexity from O($n^2$) to O(n). FastBFT arranges nodes in a tree topology, and imposes that  inter-server communication and message aggregation take place along edges of the tree.
Scalable Byzantine Reliable Broadcast~\cite{scal_brb_19} presents a probabilistic gossip-based Byzantine reliable broadcast algorithm that uses samples as a replacement for quorums, and introduces threshold contagion model of message propagation.
In contrast to these work, our work emphasized the use of relaying for control message dissemination for crash fault-tolerance consensus protocols and improves the $O(N)$ cost further, by using a simple dynamic rotated aggregation technique and operating in a deterministic fashion.

Similar to the crash fault-tolerant consensus protocols, in the BFT consensus domain, multi-leader approaches~\cite{mir_bft_19,multi_bft_gupta_19} have been presented to alleviate the bottleneck at a single-leader. Mir-BFT~\cite{mir_bft_19} generalizes PBFT and allows a set of leaders to propose request batches independently, and rotates the assignment of a partitioned request hash space to leaders. To achieve multi-leader BFT consensus, Gupta et. al. propose running several instances of consensus protocol in parallel and  present coordination-free techniques to order operations across parallel instances. Finally, work on adopting BFT consensus protocols to WAN deployments~\cite{steward_08,resilientdb_20} also make use of region based grouping and aggregation of communication.

\section{Future Work and Extensions}
\label{sec:future}

\subsection{Dynamic \& Overlapping Relay Groups}

At each round, the PigPaxos leader randomly selects a node from each relay group to carry out relay and aggregation for the group on its behalf. In the basic scheme described in Section~\ref{sec:PigPaxos}, we treated the groups as if they are static and predefined. It is possible to change the configuration of relay groups on-the-fly to improve the performance or to react quickly to crashes in some of the groups. Consider a scenario where a particular configuration of relay groups starts to experience degradation in performance. In this case, the leader may reshuffle the slower nodes into some minority relay group and improve the performance.
The leader may specify a new relay group for each round of the protocol by including future group membership information or a seed value to deterministically compute such membership in every sent message.

Another relay group optimization is to allow groups to overlap. Although this decreases efficiency due to increasing the number of messages sent and received, having alternate paths to reach nodes will improve the reliability and performance predictability in environments with link volatility and failures. With the group overlap, some follower nodes may be part of more than one relay group, creating a possibility that a node casts its vote twice. However, PigPaxos remains safe, since the leader receives the IDs of all followers acknowledging the command. This allows the leader to catch a node casting more than one vote by simply tracking the IDs of nodes that have previously voted for a particular slot, a procedure already required in regular Multi-Paxos to prevent duplicate messages from breaking the consensus.

\subsection{Improving Reads}
Reading from Paxos-backed state machines is traditionally performed in one of three different ways: reading from any node, reading from the leader node, or reading by the virtue of serializing the read operation in the log. Unfortunately, each of these methods have some drawbacks. Reading from any replica, while fast, compromises the strong consistency guarantee, since the state at a replica may be stale~\cite{retroscope}. Reading from a ``stable'' leader requires additional mechanism, such as leader leases in order to prevent "split-brain" when multiple leader candidates emerge. Finally, performing the read by serializing the read operation on the log incurs latency and throughput costs since it requires a full Paxos round to complete.
Recent work on Paxos Quorum Reads (PQR)~\cite{pqr} provides an alternative method which provides strong consistency guarantees while avoiding leases and any communication with the leader.
In PQR, reads are performed by the clients directly contacting the underutilized replicas.
A client multicasts the read request to a read quorum of replicas and waits for their replies. The read is concluded if a read quorum of replicas return the same slot number for the last accepted and applied slots of the requested data item. Otherwise, a second phase of read is required to confirm that the higher accepted slot is applied and that value is returned as the result.
Since PQR relieves the leader from serving reads, the leader can serve more writes, and throughput is increased significantly.
  
PigPaxos is especially synergistic with PQR because, due to its efficient dynamic relay based aggregation mechanism, it can compensate using a larger write quorum and allow for a small-sized read quorum. Small read quorums imply that multiple read quorums can be utilized in parallel by the clients to perform the read directly with the replicas. This helps in scaling the read throughput without bogging down the leader.

The Pig communication pattern itself can also be employed for PQR reads. A read-heavy client can benefit from employing randomly rotated pig relay nodes to form aggregation trees to perform the PQR read.
If it could be possible for the client to know about the schedule of the leader's relay nodes, it can optimize the PQR reads further by tapping into the leader's relay nodes and performing PQR using the aggregated information at the relay nodes instead of contacting the leaf replicas.

\subsection{Scaling to Large Clusters}
One direction for future work is to design and evaluate further optimization opportunities in the communication tree of PigPaxos, and to scale the protocol to hundreds of nodes. This is likely to open many new challenges. When scaled to two-orders of magnitude, we can argue that the transformation is not incremental/evolutionary and would necessitate new supplementary approaches and services for coping with this scale.
For example, in the Paxos failure detector module, the threshold for a node to become a leader candidate should be modified to be conservatively high, as contesting the leader is very costly at the scale of hundred participants.
Therefore, instead of static failure detector thresholds,
new approaches are needed to set the thresholds to be proportional to the size of $N$, such that only a couple nodes at a given time period will step up to become leader candidates.

\subsection{Applying Pig to Other Protocols}
While we have applied Pig to Multi-Paxos style of protocols, it may be helpful in many other situations requiring a fan-out/fan-in style of communication at a large scale. For instance, some large planetary-scale and edge systems, such as WPaxos~\cite{wpaxos} and DPaxos~\cite{dpaxos}, perform phase-1 of Paxos across a large set of nodes for initial data placement or ownership changes. If such phase-1 occurs frequently, for example, during data ingestion or migration between regions, it may put more load on the servers, impacting both the phase-1 operation and regular replication. Applying Pig in this situation can help alleviate the issues and improve the performance of certain workloads in such large geo-scale deployments. 

Another future work item is to apply and integrate our approach for achieving scalability to modern PBFT~\cite{pbft} descendant byzantine fault-tolerant blockchain protocols, such as Tendermint~\cite{tendermint}, Casper~\cite{casper}, and LibraBFT~\cite{hotstuff,libraBFT}.

\section{Concluding Remarks}
\label{sec:conclusion}

To address the communication bottlenecks at strong consistency replication protocols, we presented the Pig communication primitive. Pig manages to reduce the leader's communication bottleneck by decoupling the decision-making at the leader from the communication at the leader. Although aggregation approaches have been known and employed in the context of weak-consistency replication protocols, Pig shows how they can be effectively applied to and integrated with the strong consistency distributed consensus protocols, such as Paxos.

Our evaluations show that PigPaxos can provide more than 3 folds improvement in throughput over Paxos and EPaxos with little latency deterioration. Our experiments also show that the aggregation used in PigPaxos has little latency overhead as compared to the latency of Paxos.
The communication piggybacking and aggregation technique used in PigPaxos is readily applicable to many Paxos implementations~\cite{raft,etcd} and variants~\cite{fpaxos,wpaxos}, and we conjecture that PigPaxos would be useful for implementing geo-replicated distributed databases with tens of replicas distributed over many regions around the globe.


\bibliographystyle{abbrv}
\bibliography{ac,murat,ailidani}

\end{document}